\documentclass[useAMS,usenatbib]{mn2e}
\usepackage{graphicx}
\usepackage{url}
\usepackage{indentfirst}
\usepackage{amsmath}
\newcommand{\aap}{A\&A}

\newcommand{\apj}{ApJ}
\newcommand{\apjl}{ApJL}
\newcommand{\apjs}{ApJS}
\newcommand{\apss}{Ap\&SS}

\newcommand{\mnras}{MNRAS}

\newcommand{\prd}{Phys. Rev. D}

\newcommand{\physrep}{PhR}

\newcommand{\sovat}{SvA}
\newcommand{\azh}{AZh}
\newcommand{\lsim} {\mathrel{\hbox{\rlap{\lower.55ex \hbox{$\sim$}}
                             \kern-.3em \raise.4ex \hbox{$<$}}}}

\newcommand{\beq}{\begin{equation}} 
\newcommand{\eeq}{\end{equation}}
\newcommand{\be}{\begin{eqnarray}}
\newcommand{\ee}{\end{eqnarray}}

\def\lsim{\,\lower2truept\hbox{${< \atop\hbox{\raise4truept\hbox{$\sim$}}}$}\,}
\def\gsim{\,\lower2truept\hbox{${> \atop\hbox{\raise4truept\hbox{$\sim$}}}$}\,}

\title{Semi-analytical description of clumping factor and CMB free-free distortions from reionization}

\author[Trombetti T. and Burigana C.]{T. Trombetti$^{1}$ and C. Burigana$^{1,2}$ \\ \\
$^1$INAF-IASF Bologna, Via Piero Gobetti 101, I-40129, Bologna, Italy \\
$^2$Dipartimento di Fisica e Scienze della Terra, Universit\`a degli Studi di Ferrara, Via Giuseppe Saragat 1, I-44100 Ferrara, Italy}

\begin{document} 
\date{Draft version \today} 
 
 
\maketitle 
 
\label{firstpage} 
 
%%%%%%%%%%%%%%%%%%%%%%%%%%%%%%
\begin{abstract} 
%%%%%%%%%%%%%%%%%%%%%%%%%%%%%%

The density contrast of the Universe, parametrized in terms of the matter power spectrum and its variance, can amplify the signal of the free-free process in the plasma. 
The damping of fluctuations on scales smaller than the DM particle free streaming scale corresponds to a suppression of the total matter power spectrum on large wave numbers $k$.
We derive the time evolution of the variance of the matter power spectrum for various cosmological models and parameters
by numerically computing the power spectrum with a modified version of the Boltzmann code CAMB, for different values of the cut-off parameter $k_{max}$. 
Suitable analytical approximations of the numerical results are presented.   
We then characterize the CMB free-free spectral distortion accounting for the amplification effect coming from clumping factor. Indeed, the clumpiness, associated to the density contrast of the intergalactic medium,
increases at decreasing redshift. The analysis is carried out for selected astrophysical and phenomenological cosmological reionization histories, for which we evaluate the impact of the clumping factor on the free-free distortion and discuss the wavelength dependence of the predicted signal. Finally, we address a comparison with other classes of unavoidable CMB spectral distortions and future observational perspectives.
While Comptonization from reionization is dominant at high frequencies, the free-free signal predicted in the considered models contributes to the distortion at a level of few (few tens) per cent 
at frequencies below $\sim 25$\,GHz ($\sim 10$\,GHz) and represents the main signature below $\sim 4$\,GHz. The cosmological signal from the HI 21-cm background is found to prevail over the 
free-free distortion in a restricted, model dependent frequency window between $\sim 0.1$ and $\sim 0.2$\,GHz.

\end{abstract}
%
%\keywords{Cosmology: cosmic microwave background, polarization anisotropies, cosmological reionization} %\arxivnumber{1234.5678}
%%%%%%%%%%%%%%%%%%%%%%%%%%%%%%%%%%%%%%%%%%%%%%%%%%%%

\section{Introduction}

After the electron-pair annihilation, 
the evolution of the photon occupation number of cosmic microwave background (CMB) photons
in the cosmic plasma is suitably described by the Kompaneets equation \citep{Kompaneets56}. This approximation of the generalized kinetic equation includes, in its generalized form, Compton scattering, that conserves the photon number, and all the possible photon emission or absorption processes, such as (at least) double Compton and bremsstrahlung. Compton and double Compton processes depend linearly on the baryon density in the Universe, while the bremsstrahlung term is proportional to the square of the baryon density. After the recombination epoch, an accurate description of this process should account for an inhomogeneous, evolving intergalactic medium (IGM). Indeed, the matter power spectrum depends on cosmological parameters and, obviously, it is a function of the wavenumber which is inversely proportional to the linear scale. The so-called matter transfer function, $T(k)$, defines the matter power spectrum evolution from a primordial time, usually identified at the end of the inflationary stage, to a desired time, and is associated to the 
growth of the perturbations. Therefore,
the matter density contrast of the Universe, related to the evolution of the matter power spectrum, amplifies the signal of the free-free process in the plasma with respect to the case of a uniform medium. The analysis presented in this work is carried out in the context of astrophysical and phenomenological cosmological reionization histories inferred from recent astronomical and cosmological data.

In this paper, we present numerical results and analytical approximations for the amplification factor of the free-free rate and distortion parameter
derived exploiting the cosmological Boltzmann code CAMB\footnote{http://camb.info/}, properly modified to ingest different reionization histories.
This investigation, although originally designed to describe the free-free spectral distortion related to an inhomogeneous IGM,
can be applied to other studies that may need an accurate estimation of the IGM density contrast itself.

In Sect. \ref{theory} we introduce the fundamental concepts to characterize the free-free rate in an inhomogeneous medium (see also Appendix \ref{sahaEq}). Sect. \ref{clumpfact} is devoted to the computation of the clumping factor in different cosmological models and to the derivation of suitable approximations of its time evolution (see also Appendix \ref{impleprod}). In Sect. \ref{FF} we combine the recipes of previous sections to compute the CMB free-free distortion in the considered reionization scenarios. The main results are presented in Sect. \ref{results} (see also Appendix \ref{ybvalues}). Finally, we draw our main conclusions in Sect. \ref{conclu} and 
discuss our main results and the observational perspectives at radio to sub-millimeter wavelengths in Sect. \ref{discussion}.
 
\section{Theoretical framework}
\label{theory}
 
During the interaction between radiation and ionized matter, the photon equilibrium distribution function is described by the Planck law, while, more in general, the evolution in time of the photon occupation number, $\eta(x,t)$,  is represented by the complete kinetic equation \citep{Zeldovich69, 1975SvA....18..691I, DaneseZotti, BuriDanZotAA91, Hu93}: 

\begin{eqnarray}
\frac{\partial \eta(x,t)}{\partial t} & = &  \left( \frac{\partial \eta}{\partial t} \right)_{C} + \left( \frac{\partial \eta}{\partial t} \right)_{B} + \left( \frac{\partial \eta}{\partial t} \right)_{DC}\nonumber \\ & + & \left( \frac{\partial \eta}{\partial t} \right)_{cyc} + \left( \frac{\partial \eta}{\partial t} \right)_{sources},
\label{ref:boltz}
\end{eqnarray} 

\noindent where the first term in the right hand side is the collision term, hence the Compton scattering (C), and the other terms refer to photon sources, bremsstrahlung (B),
double or radiative Compton (DC) \citep{Lightman81,1984ApJ...285..275G}, cyclotron process (cyc) \citep{ZizzoBuri05} and other possible photon production processes. A dimensionless, redshift invariant frequency \citep{BuriZotDan95}, $x = h \nu / k T_{0}$, is typically adopted for numerically solving this equation. Here $aT_{0}^{4}$ defines the current CMB radiation energy density independently of the CMB spectrum shape, $T_{0} = 2.725$\,K \citep{mather99} being usually referred as the present radiation temperature.

\indent The Kompaneets equation, a convenient approximation of the more general kinetic equation, describes the evolution of the photon occupation number. Considering Compton scattering and bremsstrahlung, it can be expressed by:

\begin{eqnarray}
%\frac{\partial \eta}{\partial t} & = & \frac{1}{t_{C} x^{2} \phi} \frac{ \partial }{\partial x} \left [ x^{4} \left [ \phi \frac{\partial \eta}{\partial x} + \eta (1 + \eta) \right ] \right ] \nonumber \\
%& + & \left [ K_{B} \frac{g_{B}}{x_{e}^{3}} e^{-x_{e}} + K_{DC} \frac{g_{DC}}{x_{e}^{3}} \right ] \left [ 1- \eta (e^{x_{e}} - 1) \right ]
\frac{\partial \eta}{\partial t} & = & \frac{1}{t_{C} x^{2} \phi} \frac{ \partial }{\partial x} \left [ x^{4} \left [ \phi \frac{\partial \eta}{\partial x} + \eta (1 + \eta) \right ] \right ] \nonumber \\
& + & K_{B} \frac{g_{B}}{x_{e}^{3}} e^{-x_{e}}  \left [ 1- \eta (e^{x_{e}} - 1) \right ]
\label{deeta}
\end{eqnarray} 

\noindent where $g_{B}$ 
%and $g_{DC}$ are the
is the Gaunt factor and $x_{e} = x ( T_{e}) = x/ \phi$, being $T_{e}$ and $T_{r} = T_{0} (1 + z)$ the electron and radiation temperature, respectively, $\phi = T_{e}/T_{r}$ and $z$ the redshift. The coefficient $K_{B}(z)$ is given by: 

\begin{equation}
K_{B}(z) = \frac{8 \pi}{3} \frac{e^{6}h^{2} n_{e}(n_{H} + 4n_{He})}{m(6 \pi mkT_{e})^{1/2}(k T_{e})^{3}} = K_{0B}(z) \phi^{-7/2} \; ,
\label{ref:kappa}
\end{equation} 

\noindent being 

\begin{equation}
K_{0B}  \simeq 2.6 \cdot 10^{-25} \left ( \frac{T_{0}}{2.7 K} \right )^{-7/2} (1 + z)^{5/2} \hat{\Omega}_{b}^{2} \hspace{3 mm} \rm{sec^{-1}} \; .
\label{ref:kappa0}
\end{equation} 

%\noindent and 

%\begin{eqnarray}
%K_{DC} ( z) & = & \frac{4 \alpha}{3 \pi t_{\gamma e}} \left ( \frac{k T_{e}}{m c^{2}} \right )^{2} \int_{0}^{\infty} (1 + \eta) \eta x_{e}^{4} d x_{e} \nonumber \\ & \simeq & 8.15 \cdot 10^{-40} \left ( \frac{T_{0}}{2.7 K} \right )^{2} \; ,
%\end{eqnarray} 

\noindent
%$\alpha$ is the fine structure constant, 
In Eq. (\ref{deeta}), the kinetic equilibrium timescale between matter and radiation is expressed by: 

\begin{eqnarray}
t_{C} & = & t_{\gamma e} \frac{m c^{2}}{k T_{e}} \nonumber \\
& \simeq & 4.5 \cdot 10^{28} \left ( \frac{T_{0}}{2.7 K} \right )^{-1} \cdot \phi^{-1} \hat{\Omega}_{b}^{-1} (1 + z)^{-4} \hspace{3 mm} \rm{sec}  \; , 
\end{eqnarray}

\noindent
being $t_{\gamma e} = (n_{e} \sigma_{T} c)^{-1}$ the photon electron collision time.

\noindent
Here, the term $\hat{\Omega}_{b}$, related to the baryon density in units of the critical density and to the Hubble constant $H_{0}$, is defined as 
$\hat{\Omega}_{b} = \Omega_{b} \left [{H_{0}}/({50\, \rm{Km\, s^{-1} Mpc^{-1}})} \right]^{2}$.
Eqs.\; (\ref{ref:kappa}) and (\ref{ref:kappa0}) point out the proportionality of the bremsstrahlung term to the square of the baryon density. This makes necessary to take into account the density contrast in the intergalactic medium.

\subsection{Expansion time}

The evolution of the background quantities of the Universe depends on the contributions of the different types of energy densities. After a radiation or, more in general, relativistic particle energy density dominated phase of the Universe, (cold) matter energy density starts to dominate at $z \lsim 3000$, followed, at $z$ less than few units, by a dark energy or cosmological constant dominated epoch.
Considering a cosmic scale factor, $\omega = m_{e} c^{2} / k_{B}T_{r}$, i.e. normalized when the CMB temperature was $k_{B} T_{r} = m_{e} c^{2}$ \citep{silkstebbins83}, and taking into account the recent acceleration of the Universe, parametrized by the cosmological constant $\Omega_{\Lambda}$,
and the curvature term $\Omega_{K}$, the expansion of the Universe is governed by the equation for $\dot{\omega}$ \citep{proburi09}:

\begin{eqnarray}
\frac{1}{\dot \omega} & = & \frac{1}{d\omega/dt} \nonumber \\
& = & \frac{\tau_{g1} \omega}{\left [1+\beta \omega \left ( 1 + \frac{(\Omega_{K} / \Omega_{m}) \omega}{2.164 \cdot 10^{9}} + \frac{(\Omega_{\Lambda} / \Omega_{m}) \omega^{3}}{(2.164 \cdot 10^{9})^{3}}  \right ) \right ]^{1/2}} \; ,
\end{eqnarray}

\noindent where $\beta$ is the initial ratio between matter and photon energy densities and $\tau_{g1}$ can be seen as an initial gravitational time scale, defined as:

\begin{eqnarray}
\beta & = & \frac{\rho_{m1}}{\rho_{r1}} = 3.5 \cdot 10^{-6} h_{50}^{2} \Omega_{tot} (T_{0}/2.7 {\rm K})^{-3} \; ,\\
\frac{1}{\tau_{g1}} & = & \left ( \frac{8\pi}{3} G \rho_{r1} \right )^{1/2} = \left [ \frac{8\pi}{3} G \frac{a}{c^{2}} \left ( \frac{m_{e}c^{2}}{k} \right )^{4} \right ]^{1/2} \nonumber \\ & = & 0.0763 \;  {\rm s}^{-1} \: .
\end{eqnarray}

Accounting also for the relativistic neutrinos contribution to the Universe's dynamic, the above gravitational time scale term becomes
%$ \tau_{g1} = 13.11022 / (\kappa_{\nu})^{1/2} $
$ \tau_{g1} = 13.11 / (\kappa_{\nu})^{1/2} $, where $\kappa_{\nu}$ is the present ratio of neutrino to photon energy densities:

\begin{equation}
\kappa_{\nu} = \frac{1}{2} \left ( g_{\gamma} +  \frac{7}{8} g_{\nu} N_{\nu} (N_{\nu}^{eff})^{(-4/3)} \right) = 1 + \frac{7}{8} N_{\nu} \left ( \frac{4}{11} \right )^{4/3} \; ,
\end{equation}

\noindent being $g_{\gamma}$ the effective number of photons spin states, $N_{\nu}$ the number of relativistic, 2-component neutrino species, $N_{\nu}^{eff}$ the effective number of species at the decoupling epoch. Typically, for 3 species of massless neutrinos and for 2 massless photon spin states, $\kappa_{\nu} \simeq 1.68$.

\subsection{Bremsstrahlung process}
\label{brem}

Defining the density as the contribution of a mean term plus a small perturbation, $\rho = \langle \rho \rangle(1 + \delta \rho /  \langle \rho \rangle)$, and averaging over a representative volume of the Universe
%Defining the density as the contribution of a mean term plus a small perturbation, $\rho = \langle \rho \rangle(1 + \delta)$, and averaging over a representative volume of the Universe

%\begin{equation}
%\frac{\langle \rho^{2} \rangle }{\langle \rho \rangle^{2}} =  1 + \langle \delta^{2} \rangle = 1 + \sigma^{2} > 1 \, .
%\end{equation} 
\begin{equation}
\frac{\langle \rho^{2} \rangle }{\langle \rho \rangle^{2}} =  1 + { \langle (\delta \rho)^{2} \rangle \over \langle \rho \rangle^{2} }= 1 + \sigma^{2} > 1 \, .
\end{equation} 

This defines the clumping factor, that, in this context, is a time dependent multiplicative factor to be accounted to modify the bremsstrahlung rate from the case of homogeneous medium to the case of inhomogeneous medium.
In Sect. \ref{clumpfact} we will derive numerical and semi-analytical estimates of $1 + \sigma^{2}$ for a set of different cosmological models.
We then get the complete free-free term in an inhomogeneous medium multiplying the usual baryonic density parameter in Eq. (\ref{ref:kappa0})
by this factor, i.e.

\begin{equation}
\Omega_{b}^{2} (z) = \Omega_{b,homog}^{2}  (1+\sigma^{2}(z)) \; ,
\end{equation}

\noindent where $\Omega_{b,homog}^{2}$ corresponds to the (standard) homogeneous case. Note that the density contrasts of free electrons and atoms in different ionization states could be in principle different. 
Also, inhomogeneities in the medium temperature could be in principle taken into account. On the other hand, being weak the dependence on temperature of 
$K_{B}/x_{e}^{3} \propto \phi^{-1/2}$ and of the term $e^{-x_{e}} \simeq 1-x_{e}$ at low frequencies, where free-free process is particularly relevant, we can neglect to first order 
this further correction.
Higher order corrections could be searched evaluating the variances of the various terms in Eq. (\ref{ref:kappa}). Although an accurate analysis of these effects is out of the scope of this paper, 
in Sect. \ref{discussion} we will present some numerical estimates based on currently available studies on IGM temperature.

The separation approach exploited in this work, based on the factorization of the square of the baryon density, allows to get the main effect from matter density contrast and to couple it with all the other relevant aspects in the thermal and ionization history, evaluated considering averaged functions, with versatility regarding the underlying cosmological model. 

The total Gaunt factor (see \cite{RybLight79}) appearing in Eq. (\ref{deeta}), $g_{B} (x_{e})$, has been evaluated following the simple formulas in \cite{BuriZotDan95} (see 
\cite{itoh00} 
for higher accuracy expressions) that, when properly account for the ion specie, approximate to per cent level the results by \cite{KarzasLatter}, an accuracy compatible with, or better than, the precision coming from currently unavoidable uncertainties in astrophysical models considered in this work. A few implementation allows to ingest the adopted fractions of hydrogen and helium. Indeed, the free-free rate defined in Eq. (\ref{ref:kappa}) holds in a fully ionized medium. Actually, being this analysis focused at intermediate and low redshifts, resulting into late spectral distortions, this condition is not reasonable at all since the plasma, during this period, stands in different ionization states. The initial heating redshift is set in fact in the code at $z_{h} = 30$. Precisely, the electrons, hydrogen and helium number densities become $n_{e}(n_{H} + 4n_{He}) = n_{e}^{F}(n_{H^{+}} + 4n_{He^{++}} +n_{He^{+}})$, being the free electron fraction $n_{e}^{F} = \chi_{e}n_{e}$, with $n_{e}$ the total electron fraction, 
$n_{e} = n_{b} (1+f_{H}) / 2 $, and where $n_{b}$ is the baryon number density, provided by:

\begin{equation}
n_{b} = \frac{\rho_{b}}{m_{b}} = 2.8 \cdot 10^{-6} \hat{\Omega}_{b} (1+z)^{3} \; .
\end{equation}

The ionization fraction, $\chi_{e}$, is determined by the reionization history. The repartition of atoms in different ionization states has been evaluated, at equilibrium, by solving the Saha equations, a system that describes the ratio between two different ionization states of an element (see Appendix \ref{sahaEq}). The weighted total Gaunt factor is:

\begin{equation}
g_{B} (x_{e}) = (\chi_{H^{+}} + \chi_{He^{+}})g_{B}^{H} (x_{e}) + \chi_{He^{+}}g_{B}^{He} (x_{e})\; .
\end{equation}

The photon occupation number at a time $t$, assuming an initial blackbody spectrum, $\eta_{BB,i}$, under the only effect of bremsstrahlung as photon emission-absorption process can be well represented 
by the relation \citep{BuriZotDan95}:

\begin{equation}
\eta \simeq \eta_{BB,i} + \frac{y_{B}}{x^{3}} - u\frac{2}{x/ \phi} \, ,
\label{appsol}
\end{equation} 

\noindent with $u=u(t)$ the Comptonization parameter, and $\phi_{i}$ the initial electron temperature necessary for the evaluation of the distortion parameter, related to the fractional amount of energy exchanged between matter and radiation by $\phi_{i} = T_{i}/ T_{r} =  (1 + \Delta \varepsilon / \varepsilon_{i})^{-1/4} \simeq 1 - u$ (where $u$ is here computed at the final epoch).

Note that we are interested here at wavelengths larger than few millimeters where the Comptonization spectral shape is well approximated by the usual brightness temperature decrement characterizing the Rayleigh-Jeans region. At shorter wavelengths it should be replaced by the complete Comptonization formula.

At very low frequencies, the free-free distortion becomes self-absorbed. A formal, complete solution in the case of small late distortions under the combined effect of Compton scattering and bremsstrahlung 
\citep{1980A&A....84..364D}, including spontaneous and induced emission and absorption, can be found in sect. 3.3 of \cite{BuriZotDan95}. On the other hand, in the frequency range $x_{B} \ll x \ll 1$ (being $x_{B}$ the frequency at which the Universe optical depth for free-free absorption, $y_{abs, B}$ \citep{BuriZotDan95}, is unit \citep{dezotti1986}) the expression in 
Eq. (\ref{appsol}) very well approximates the complete solution. 
The validity of the assumption $x \gg x_{B}$ for the range of wavelengths considered in this work has been numerically checked for the suppression model at $k_{max}$ = 100 including the details of ionization and thermal histories, but it is certainly satisfied by other cutoff values and reionization histories, since they exhibit not so different clumping factors and free-free rates. 
In particular, we find $y_{abs,B} = 1.157 \cdot 10^{-22}$ at $\lambda$ =\,0.01\,cm and, obviously, a much greater value $y_{abs,B} = 1.947 \cdot 10^{-9}$ at $\lambda$ =\,10\,m.

Finally, the free-free parameter $y_{B} (t)$ turns out to be:

\begin{eqnarray}
y_{B} & = & \int_{t_{h}}^{t} (\phi - \phi_{i}) \phi^{-3/2} g_{B} (x, \phi) K_{0B} dt \nonumber \\ & = & \int_{z}^{z_{h}} (\phi - \phi_{i}) \phi^{-3/2} g_{B} (x, \phi) K_{0B} t_{exp} \frac{dz}{1+z} \, .
\label{eq:free}
\end{eqnarray} 

\subsection{Cold and Warm Dark Matter models}
\label{matpow}

The matter density contrast depends on the cosmological model and parameters. In particular, the nature of dark matter particles affects the power spectrum of density perturbations at small scales, with implications for 
the clumping factor.

The Cold Dark Matter (CDM) standard cosmological model forecasts the existence of primordial cold and collisionless particles with negligible small velocity dispersion at the epoch of radiation-matter equality and a derived matter power spectrum supporting small scale structure formation. The dark matter velocity distribution suppresses fluctuations below its free streaming scale, proportional to the mean velocities-mass ratio. In this framework, galaxy formation is a hierarchical process, resulting in the typical {\it bottom up} scenario where denser clumps give rise to satellite galaxies \citep{Boyanovsky}, leading to the well known over-prediction of these galaxies with respect to those observed in the Milky Way.

In this context, the idea of Warm Dark Matter (WDM) particles was proposed as a possible solution of the small scales CDM and cuspy halo problems, the latter related to the density profile in virialized DM halos. 

WDM candidates have intrinsic thermal velocities, so exhibit larger velocity dispersions, compared to CDM particles, and present characteristic free streaming scales, which contribute to determine clustering properties, such that small scale structures formation is suppressed.  

Typical particle masses are in the range of the keV scale, which is intermediate between cold and hot DM masses, $m_{warm} \sim (1-10)$\,keV \citep{Sanchez}, being $m_{cold} \sim (10-10^{2})$\,GeV and 
$m_{hot} \sim$ few eV. Sterile neutrinos or gravitinos are possible candidates for WDM.

The suppression of fluctuations due do $\Lambda$WDM particles on scales smaller than their free-streaming scale, corresponding to a cut-off of the total matter power at large $k$, slows down the growth of structure 
\citep{Viel} and can be described by means of the transfer function for different cut-off values. 

For a given cosmological model, the transfer function describes the effect induced by the free-streaming length on the matter distribution such:

\begin{equation}
T(k) = \left [ \frac{P(k)_{WDM}}{P(k)_{CDM}}\right ]^{1/2}.
\end{equation} 

While the formation of the first generation of stars is affected by the adopted cosmological model, large scale structure distributions are independent of the model \citep{Gao}. Thus, in this analysis we assumed a CDM model with a cut-off $k_{max}$ in the range (20,1175) which approximately mimics the power spectrum drop for WDM models with different particle properties. 

\subsection{Reionization histories}

The epoch of reionization, driven by the formation of the first luminous objects at $z \sim 7-12$, as probed by the discovery of galaxies and quasars at $z \gsim 6$ and by the Gunn-Peterson test \citep{Fan2006}, is one of the main tool to unravel the initial stages of structure formation. During this phase, ionizing radiation was transferred to the inhomogeneous cosmic gas in several distinct steps. The physical properties of the astrophysical sources responsible for this process largely affect the evolution of the medium in the reionization era \citep{BarkLo01}. In particular, during the hydrogen reionization phase it is possible to identify an initial pre-overlap stage and an intermediate and quick enough overlap period which ends up in the reionization process. The same ionizing sources should induce the first ionization of helium, while the second one should be delayed to smaller redshift, being therefore in principle accessible to observations (through Ly-$\alpha$ absorption lines). 

The intergalactic ionizing radiation field is typically characterized by the so called escape fraction, the amount of ionizing radiation supplied by stars and quasars, which depends on the radiative efficiency of these sources, their emission spectrum, and their density distribution. In principle, it is possible to distinguish two contributions, one coming from galactic gas and one from galactic dust, both of them able to absorb and re-emit electromagnetic radiation at lower frequencies \citep{Benson00}. Hence, the global escape fraction is strictly related to galaxy mass and redshift, so it should be tracked within the densest regions in the Universe, where the transfer of ionizing radiation is more efficient. 

Usually, reionization era is associate to the Thomson optical depth, $\tau$, an integrated cosmological parameter which conveys information on the physical conditions of the Universe due to recent events, opening an observational window on the physics of ancient cosmic objects. Furthermore, it is clearly linked to later non linear evolution.

In order to investigate the effects triggered by the reionization process and amplified by a non negligible IGM density contrast on CMB free-free spectral distortions, we focalized our analysis on three well determined Universe's reionization models. Two of them, namely suppression (S) \citep{Choudhury2006} and filtering (F) \citep{Gnedin2006}, are astrophysical motivated, the other is a phenomenological, late double peaked (L) history \citep{pavel}, considered here for comparison 
(see \cite{trombetti_burigana_JMP} and references therein for further details about these models and their parameters). 
The S and F reionization scenarios can be tuned to match various kinds of astronomical observations (see \cite{paperI}). Their optical depths are then well determined: 
$\tau_{S} = 0.1017$ and $\tau_{F} = 0.0631$, respectively.
The L scenario can be modeled by means of free quantities describing the electron ionization fraction and electron temperature. We set its main parameters in order to have $\tau_{L} = \tau_{S}$.

Finally, we exploited also the reionization history produced with the standard CAMB assuming the cosmological parameters derived by the {\it Planck} Collaboration\footnote{www.rssd.esa.int/Planck}
\citep{2013arXiv1303.5076P}, presented with the recent first release of cosmological data.

\section{Semi-analytical description of clumping factor}
\label{clumpfact}

The clumping factor, related to the density of the IGM, affects the escape of radiation from an inhomogeneous medium \citep{BarkLo01}. Typically, assuming an uniform ionized medium, the clumping quantifies the number of spatially averaged recombinations in the IGM relative to the ones in the cosmic gas, per time and volume unit \citep{PawlikAl09}. Commonly, the clumping is supposed to be roughly spatially uniform providing a ionized volume much greater than the clumping scale. This parameter also has an impact on the gas recombination timescale, such that the higher is the clumping value, the faster is the recombination process.

Free-free emission from the ionized medium, with the interplay of cosmic radiative feedback of the sources responsible for Universe reionization, can produce a potentially remarkable distortion in the CMB spectrum, particularly at long wavelengths. 
\cite{Salvaterra09} estimated lower limits of this effect for the filtering and the suppression model explicitly assuming a diffuse, averaged density, i.e. neglecting clumping, but suggesting the relevance of including it in the computation. 

In order to derive the variance of the matter power spectrum we run CAMB, the code for the anisotropies in the CMB. It allows in fact a precise computation of the power spectrum of all the relevant components of the cosmic fluid at the desired grid of redshifts. The variance of the baryonic matter can be then calculated integrating over a suitable wavenumber interval the corresponding power spectrum:

\begin{equation} 
\sigma^{2} (z) = {1 \over 2\pi^{2}} \int P(k, z) k^{2 }dk \, .
\end{equation} 

Clearly, the most accurate choice is to perform this computation for the interesting set of cosmological parameters for the considered models. On the other hand, this is highly demanding in terms of computational time (and, possibly, solution storage) for the exploration of a wide set of cosmological models. 

In this section we provide a set of suitable approximations able to represent $\sigma^{2} (z)$ at an accuracy level of a few per cent for the models considered here, exploiting also changes in two parameters crucial in this context, the amplitude and the small scale cut-off of perturbations, and some other cosmological models. The found formulas can be also adopted as guidelines to derive approximations valid for other models.

In the context of this work, this analysis represents an intermediate step for the evaluation of the free-free distortion, but, more in general, the found results can be useful to improve the treatment of clumping factor
in other cosmological applications, as, for example, the study of alternative reionization scenarios (see e.g. \citet{salvaterra_etal_11}).

\subsection{Suppression model}
\label{clump_supp}

We start from the cosmological parameters adopted in the suppression reionization history (see \cite{Salvaterra09}). With reference to CAMB notation, they are:

\begin{equation}
\begin{tabular}{ l l l  }
$\tau = 0.1017$ & $n_{s} = 0.95$ & $\sigma_{8} = 0.74$ \\
$\Omega_{b} = 0.0413$ & $\frac{d n_{s}}{d \ln k} = 0$ & $Amp = 2.018\cdot 10^{-9}$ \\
$\Omega_{cdm} = 0.1987$ & $r = 0.1$ & $z_{reion} = 10$ \\
$\Omega_{\Lambda} = 0.76$ & $h = 73$ & $w = -1$ \\
\end{tabular}
\end{equation} \\

The CAMB code allows us to save a file of the evaluated matter power spectrum in $h$/Mpc units, normalized for baryons, cold dark matter particles and massive neutrinos, and the transfer function in the synchronous gauge, given a unit primordial curvature perturbation on superhorizon scales, for each requested redshift and for each parameter that represents the cut-off in the estimate of the variance, the transfer $k$ maximum. We selected a redshift interval between $0$ and $30$ with an increasing step of $0.1$, and an initial $transfer\_kmax=1175$ (hereafter $k_{max}$), to be then able to truncate the integral at the desired wavenumber.

For the computation of the variance we used the routine $D01GAF$ of the NAG libraries that evaluates the integral of a function (defined numerically at four or more points) over the whole specified range, using third order finite difference formulae with error estimates, according to a method discussed by \cite{GillMiller}. 

For this purpose, we implemented a Fortran routine which derives the variance for the above set of redshifts and stores it in a file. At each run we adopted different values of the input parameter $k_{max}$ in the NAG routine, such as $20$, $50$, $100$, $150$, $200$, $250$, $300$, $350$, $400$, $500$, $600$, $700$, $800$, $850$, $900$, $1000$, $1175$, to explore the differences in the shape of the curves, and looking for an analytical description of the results valid for the whole set of parameters. With this aim, we adopted the case $k_{max} = k_{ref} = 100$ as a reference. We found for it an analytical relation with the fitting functions of the program {\it Igor Pro} and searched for a general function able to reproduce the other curves of variance just depending on the variation of the cut-off parameter.

\begin{figure}
\centering
\includegraphics[scale=0.5]{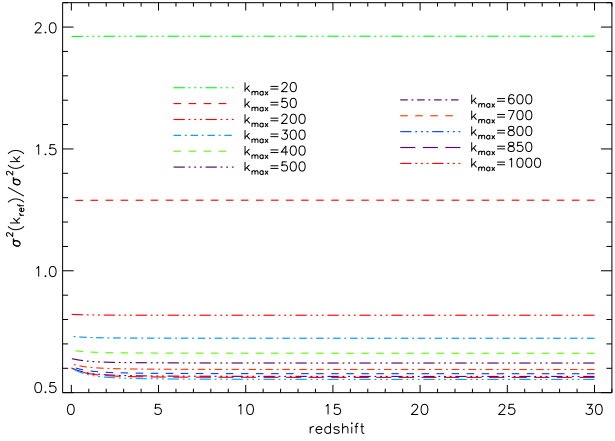}
\caption{Linear dependence of the ratio between the variance at a generic cut-off and the variance of the reference curve with $k_{ref} = 100$.}
\label{fig:lin}
\end{figure}

In all cases, the ratio between the variance's reference case and the one at a generic cut-off was best fitted quite well by a linear function of the redshift, for which the intercept and the slope, say $a(k)$ and $b(k)$, depends only on the chosen $k_{max}$, as shown in Fig.\;\ref{fig:lin}.

To generalize also this behavior, we looked for an universal law for these parameters, able to reproduce the results for any other cut-off's value $k = k_{max}$.  The two new parametric quantities, $a(k)$ and $b(k)$, were extremely good represented in terms of the function Double Exponential X Offset (see Appendix \ref{impleprod} for the fit coefficients): 

\begin{equation}
\begin{cases}
a(k) = a_{0} + a_{1} {\rm exp}(\frac{a_{2}-k}{a_{3}}) + a_{4} {\rm exp}(\frac{a_{2}-k}{a_{5}})\\\\
b(k) = b_{0} + b_{1} {\rm exp}(\frac{b_{2}-k}{b_{3}}) + b_{4} {\rm exp}(\frac{b_{2}-k}{b_{5}})
\end{cases} , 
\label{eq:ab}
\end{equation}

\noindent so that we could establish the relation: 

\begin{equation}
\frac{\sigma^{2}(k_{ref})}{\sigma^{2}(k)}(z) = \sigma^{2}_{ratio}(z) = a(k) + b(k) z.
\label{fitratio}
\end{equation} 

When fitting the case $k_{max}=100$, now dividing the range of redshift in two subintervals, $z_{l} \mathcal{2} [0;9]$ and $z_{h} \mathcal{2} [9;30]$, the most accurate representation is, again, in terms of the function Double Exponential X Offset (see also Appendix \ref{impleprod}):

\begin{equation}
\begin{cases}
\sigma^{2}_{fit}(k_{ref})^{low} = l_{0} + l_{1} {\rm exp}(\frac{l_{2}-z_{l}}{l_{3}}) + l_{4} {\rm exp}(\frac{l_{2}-z_{l}}{l_{5}})\\\\
\sigma^{2}_{fit}(k_{ref})^{high} = h_{0} + h_{1} {\rm exp}(\frac{h_{2}-z_{h}}{h_{3}}) + h_{4} {\rm exp}(\frac{h_{2}-z_{h}}{h_{5}})
\end{cases} ,
\label{eqfit}
\end{equation} 

\begin{figure}
\centering
\includegraphics[scale=0.59]{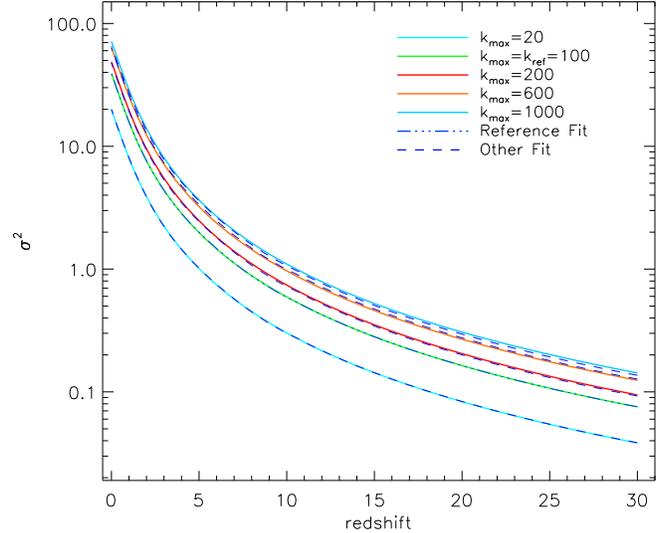}
\caption{Variance for different values of the cut-off parameter $k_{max}$ with the relative fitting functions. The curve {\it Reference Fit} is the fit to the $k_{max}=100$ curve, while {\it Fit other $k_{max}$} are the other curves obtained with analytical functions derived from the reference one as explained in the text.}
\label{fig:var}
\end{figure}

\begin{figure}
\centering
\includegraphics[scale=0.5]{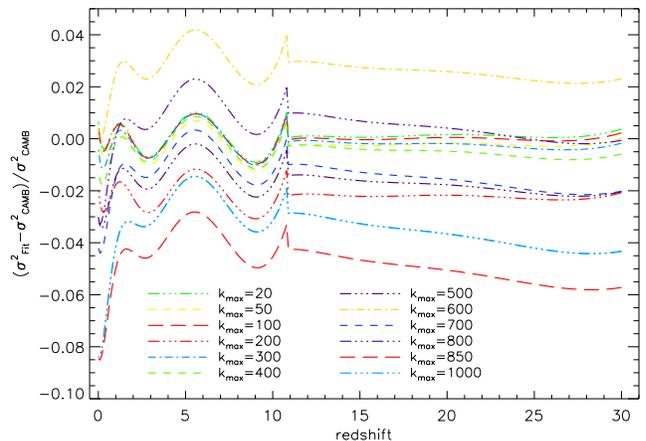}
\caption{Variance relative errors for the fitting functions of the previous plot.}
\label{fig:errsup}
\end{figure}

Finally, we extrapolated a generalized analytic relation for the evaluation of the variance for each desired cut-off parameter from the reference one, as:

\begin{equation}
\sigma^{2}_{analytic}(k) = \frac{\sigma^{2}_{fit}(k_{ref})}{\sigma^{2}_{ratio}(z)} \, ,
\label{ref:fit}
\end{equation} 

\noindent where $\sigma^{2}_{fit}(k_{ref})$ comes from the concatenation of $\sigma^{2}_{fit}(k_{ref})^{low}$ and $\sigma^{2}_{fit}(k_{ref})^{high}$. Fig.\;\ref{fig:var} shows the variance of the matter power spectrum derived from CAMB code and from this analytical fit. From the plot emerges that, to first approximation, the curves are reproduced with good accuracy also in the case of a cut-off value much greater than the reference one. The relative difference between simulated and fitted data, displayed in Fig.\;\ref{fig:errsup}, is at most of $8\%$ for the highest cut-off value\footnote{The entity of these errors are the consequence that the higher is the chosen cut-off value, the bigger is the deviation of the ratios $\sigma^{2}(k_{ref}) / \sigma^{2} (k)$ from an ideal linear profile, mostly at low redshift (see Fig.\;\ref{fig:lin} for comparison).} and typically below a few per cent.

\subsection{Exploiting other cosmological models} 

\subsubsection{Constant cut-off parameter}
\label{constcut_varcosmo}

We explore here various alternative cosmological models, looking for an analytical approximation of the variance first assuming fixed cut-off values of the matter power spectrum. To this aim, we run the original CAMB code with different combinations of cosmological parameters compatible with WMAP results\footnote{http://lambda.gsfc.nasa.gov/product/map/} \citep{larson2011,komatsu2011}.
We have taken into account two classes of theoretical models, adopting four different prescriptions for each class, whom parameter are specified in Table \ref{tableCDMhist} and Table \ref{tableWDMhist}:

1) $\Lambda$CDM + Run + Tens, the Standard Cosmological model with the inclusion of the SZ effect, a dark energy component, the tensors, the gravitational lensing and the running, the latter being the variation of the scalar spectral index with respect to the wave number $k$. 

2) wCDM, in which the dark energy equation of state is allowed to vary, hence with $w \ne -1$.

In the analysis, the cut-off parameter has been set to $k_{max}= 1175$, to allow then the truncation at the desired wavenumber.

\vskip 0.2cm

\begin{table}
\centering
\begin{tabular}{| c | c | c | c | c |}
\hline
$$ | & $\Lambda$CDM0 & $\Lambda$CDM1 & $\Lambda$CDM2 & $\Lambda$CDM3 \\
\hline
$\Omega_{b}$ & $0.046$ & $0.0427$ & $0.0509$ & $0.0487$ \\
$\Omega_{cdm}$ & $0.247$ & $0.202$ & $0.292$ & $0.229$ \\
$\Omega_{\Lambda}$ & $0.707$ & $0.7553$ & $0.6571$ & $0.7223$ \\
$\tau$ & $0.0872$ & $0.112$ & $0.104$ & $0.0966$ \\
$h$ & $69.1$ & $71$ & $66$ & $73$ \\
$n_{s}$ & $1.076$ & $1.01$ & $1.141$ & $1.05$ \\
$r$ & $0.1$ & $0.01$ & $0.03$ & $0.07$ \\
$z_{reion}$ & $11.4$ & $13.229$ & $13.094$ & $11.122$ \\
%$\frac{dn_{s}}{d\ln k}$ & $-0.048$ & $0.007$ & $0.029$ & $-0.035$ \\
$dn_{s}/d\ln k$ & $-0.048$ & $0.007$ & $0.029$ & $-0.035$ \\
$\sigma_{8}$ & $0.804$ & $0.7716$ & $0.7926$ & $0.801$ \\
$Amp$ & $1.885$ & $2.117$ & $1.582$ & $1.852$ \\
%$w$ & $-1$ & $-1$ & $-1$ & $-1$ \\
\hline
\end{tabular} 
\caption{Parameters of the $\Lambda$CDM cosmologies. In table, the amplitude ({\it Amp}) is in $10^{-9}$ units and the dark energy equation of state is constant ($w = -1$).}
\label{tableCDMhist}
\end{table}

%1a) $\Lambda$CDM0: 
%\begin{equation}
%\begin{tabular}{ l l l  }
%$\tau = 0.0872$ & $n_{s} = 1.076$ & $\sigma_{8} = 0.804$ \\
%$\Omega_{b} = 0.046$ & $\frac{d n_{s}}{d \ln k} = -0.048$ & $Amp = 1.885\cdot 10^{-9}$ \\
%$\Omega_{cdm} = 0.247$ & $r = 0.1$ & $z_{reion} = 11.4$ \\
%$\Omega_{\Lambda} = 0.707$ & $h = 69.1$ & $w = -1$ \\
%\end{tabular}
%\label{case1a}
%\end{equation} 
%
%1b) $\Lambda$CDM1: 
%\begin{equation}
%\begin{tabular}{ l l l  }
%$\tau = 0.112$ & $n_{s} = 1.01$ & $\sigma_{8} = 0.7716$ \\
%$\Omega_{b} = 0.0427$ & $\frac{d n_{s}}{d \ln k} = 0.007$ & $Amp = 2.117\cdot 10^{-9}$ \\
%$\Omega_{cdm} = 0.202$ & $r = 0.01$ & $z_{reion} = 13.229$  \\
%$\Omega_{\Lambda} = 0.7553$ & $h = 71$ & $w = -1$ \\
%\end{tabular}
%\end{equation} 
%
%1c) $\Lambda$CDM2: 
%\begin{equation}
%\begin{tabular}{ l l l }
%$\tau = 0.104$ & $n_{s} = 1.141$ & $\sigma_{8} = 0.7926$ \\
%$\Omega_{b} = 0.0509$ & $\frac{d n_{s}}{d \ln k} = 0.029$ & $Amp = 1.582\cdot 10^{-9}$ \\
%$\Omega_{cdm} = 0.292$ & $r = 0.03$ & $z_{reion} =  13.094$ \\
%$\Omega_{\Lambda} = 0.6571$ & $h = 66$ & $w = -1$ \\
%\end{tabular}
%\end{equation} 
%
%1d) $\Lambda$CDM3: 
%\begin{equation}
%\begin{tabular}{ l l l }
%$\tau = 0.0966$ & $n_{s} = 1.05$ & $\sigma_{8} = 0.801$ \\
%$\Omega_{b} = 0.0487$ & $\frac{d n_{s}}{d \ln k} = -0.035$ & $Amp = 1.852\cdot 10^{-9}$ \\
%$\Omega_{cdm} = 0.229$ & $r = 0.07$ & $z_{reion} = 11.122$ \\
%$\Omega_{\Lambda} = 0.7223$ & $h = 73$ & $w = -1$ \\
%\end{tabular}
%\end{equation} 

\vskip 0.2cm

\begin{table}
\centering
\begin{tabular}{| c | c | c | c | c |}
\hline
$$ | & $\Lambda$WDM0 & $\Lambda$WDM1 & $\Lambda$WDM2 & $\Lambda$WDM3 \\
\hline
$\Omega_{b}$ & $0.044$ & $0.022$ & $0.061$ & $0.037$ \\
$\Omega_{cdm}$ & $0.215$ & $0.137$ & $0.297$ & $0.252$ \\
$\Omega_{\Lambda}$ & $0.741$ & $0.841$ & $0.642$ & $0.711$ \\
$\tau$ & $0.088$ & $0.104$ & $0.073$ & $0.097$ \\
$h$ & $75$ & $90$ & $61$ & $78$ \\
$n_{s}$ & $0.964$ & $0.849$ & $0.98$ & $0.911$ \\
$r$ & $0.1$ & $0.05$ & $0.001$ & $0.005$ \\
$z_{reion}$ & $10.5$ & $14.486$ & $9.742$ & $13.138$ \\
%$dn_{s}/d\ln k$ & $0$ & $0$ & $0$ & $0$ \\
$\sigma_{8}$ & $0.857$ & $0.798$ & $0.81$ & $0.97$ \\
$Amp$ & $2.095$ & $1.518$ & $2.782$ & $13.138$ \\
$w$ & $-1.12$ & $-1.55$ & $-0.72$ & $-0.88$ \\
\hline
\end{tabular} 
\caption{Parameters of the $\Lambda$WDM cosmologies. In table, the amplitude ({\it Amp}) is in $10^{-9}$ units and there is no running ($d n_{s} /d\ln k = 0$). }
\label{tableWDMhist}
\end{table}

%2a) wCDM0: 
%\begin{equation}
%\begin{tabular}{ l l l }
%$\tau = 0.088$ & $n_{s} = 0.964$ & $\sigma_{8} = 0.857$ \\
%$\Omega_{b} = 0.044$ & $\frac{d n_{s}}{d \ln k} = 0$ & $Amp = 2.095\cdot 10^{-9}$ \\
%$\Omega_{cdm} = 0.215$ & $r = 0.1$ & $z_{reion} = 10.5$ \\
%$\Omega_{\Lambda} = 0.741$ & $h = 75$ & $w = -1.12$ \\
%\end{tabular}
%\end{equation} 
%
%2b) wCDM1: 
%\begin{equation}
%\begin{tabular}{ l l l }
%$\tau = 0.104$ & $n_{s} = 0.849$ & $\sigma_{8} = 0.798$ \\
%$\Omega_{b} = 0.022$ & $\frac{d n_{s}}{d \ln k} = 0$ & $Amp = 1.518\cdot 10^{-9}$ \\
%$\Omega_{cdm} = 0.137$ & $r = 0.05$ & $z_{reion} = 14.486$ \\
%$\Omega_{\Lambda} = 0.841$ & $h = 90$ & $w = -1.55$ \\
%\end{tabular}
%\end{equation} 
%
%2c) wCDM2: 
%\begin{equation}
%\begin{tabular}{ l l l }
%$\tau = 0.073$ & $n_{s} = 0.98$ & $\sigma_{8} = 0.81$ \\
%$\Omega_{b} = 0.061$ & $\frac{d n_{s}}{d \ln k} = 0$ & $Amp = 2.782\cdot 10^{-9}$ \\
%$\Omega_{cdm} = 0.297$ & $r = 0.001$ & $z_{reion} = 9.742$ \\
%$\Omega_{\Lambda} = 0.642$ & $h = 61$ & $w = -0.72$ \\
%\end{tabular}
%\end{equation} 
%
%2d) wCDM3: 
%\begin{equation}
%\begin{tabular}{ l l l }
%$\tau = 0.097$ & $n_{s} = 0.911$ & $\sigma_{8} = 0.97$ \\
%$\Omega_{b} = 0.037$ & $\frac{d n_{s}}{d \ln k} = 0$ & $Amp = 2.282\cdot 10^{-9}$ \\
%$\Omega_{cdm} = 0.252$ & $r = 0.005$ & $z_{reion} = 13.138$ \\
%$\Omega_{\Lambda} = 0.711$ & $h = 78$ & $w = -0.88$ \\
%\end{tabular}
%\end{equation} 

\begin{figure}
\centering
\includegraphics[scale=0.5]{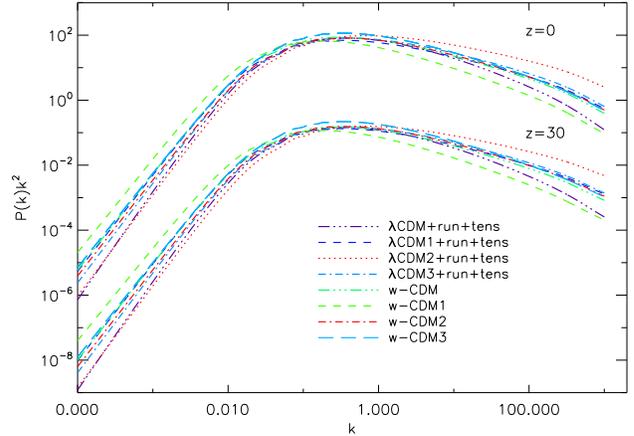}
\caption{Matter power spectrum at $z = 0$ and z = 30 for the two different cosmologies and their alternative set of parameters.} 
\label{fig:var23}
\end{figure}

As in the previous section, the variance of these models has been first evaluated with $k_{max} = 100$. The matter power spectra at $z = 0$ and $z = 30$ for the considered models
are displayed in Fig.\;\ref{fig:var23}. In both cases the trend is very similar: differences between models are more evident at small scales and tend to be smaller on intermediate scales. 

We assumed that, for the determination of the $\sigma^{2}$, we could take into account the same reference curve at $k_{max} = 100$ we already adopted in the case of the suppression model, since we are studying a generalized limiting case $k_{max} = 100$ and the coefficients $a(k)$ and $b(k)$ are $k$-independent constants.
To reproduce $\sigma^{2} (z)$ with a reasonable accuracy  in each considered model, we found that the reference curve can be simply multiplied by a model dependent correction factor, $\sigma_{corr}$, provided in Table \ref{tablesigma}: 
%
%\vskip 0.2cm
%
%$\cdot$ $\Lambda$CDM0: $ \sigma_{corr} = 1.131$,
%
%$\cdot$ $\Lambda$CDM1: $ \sigma_{corr} = 1.603$,
%
%$\cdot$ $\Lambda$CDM2: $ \sigma_{corr} = 3.354$,
%
%$\cdot$ $\Lambda$CDM3: $ \sigma_{corr} = 1.889$,
%
%$\cdot$ wCDM0: $ \sigma_{corr} = 1.447$,
%
%$\cdot$ wCDM1: $ \sigma_{corr} = 0.611$,
%
%$\cdot$ wCDM2: $ \sigma_{corr} = 1.705$,
%
%$\cdot$ wCDM3: $ \sigma_{corr} = 1.968$.
%
%\vskip 0.2cm
%

\begin{table}
\centering
\begin{tabular}{| c | c |}
\hline
Model & $\sigma_{corr}$ \\
\hline
$\Lambda$CDM$0$ & $1.131$ \\
$\Lambda$CDM$1$ & $1.603$ \\
$\Lambda$CDM$2$ & $3.354$ \\
$\Lambda$CDM$3$ & $1.889$ \\
wCDM$0$ & $1.447$ \\
wCDM$1$ & $0.611$ \\
wCDM$2$ & $1.705$ \\
wCDM$3$ & $1.968$ \\
\hline
\end{tabular} 
\caption{Parameters of the $\Lambda$WDM cosmologies. In table, the amplitude ({\it Amp}) is in $10^{-9}$ units and there is no running ($d n_{s} /d\ln k = 0$). }
\label{tablesigma}
\end{table}

\begin{figure}
\centering
\includegraphics[scale=0.5]{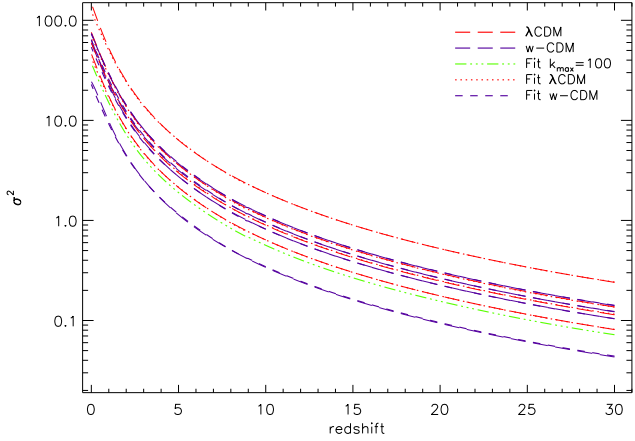}
\caption{Variance for the two classes of cosmologies and their alternative set of parameters compared to the relative fitting functions derived from the reference curve ($k_{max} = 100$).} 
\label{fig:var1}
\end{figure}

\begin{figure}
\centering
\includegraphics[scale=0.5]{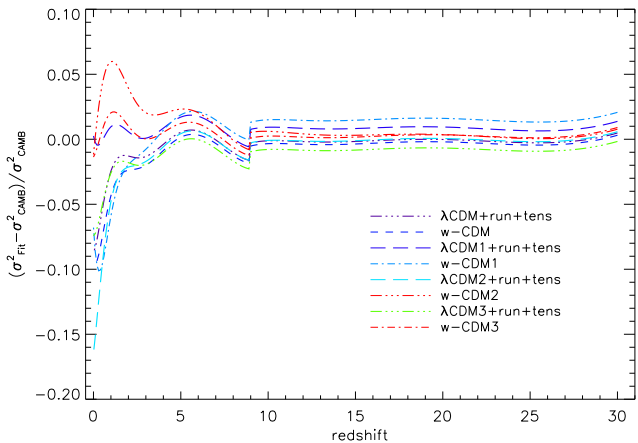}
\caption{Relative error of the variance for the models analyzed.} 
\label{fig:var4}
\end{figure}

In Fig.\;\ref{fig:var1} we show the comparison between the variance computed numerically 
and the fitting relations for the two considered cosmological models, while the relative errors of fitting formulas are reported in Fig.\;\ref{fig:var4}. 
The accuracy of the above fits is  typically better than a few per cent (except, for some models, at $z \lsim 2$) and always better than 15\%.

\vskip 0.2cm

The power spectrum shape depends on the cosmological model and the full set of parameters.
Fixing the other parameters, the perturbation amplitude mainly defines the $P(k)$ overall level. Thus, the variance scales essentially linearly with the perturbation amplitude as the power spectrum does. 
To verify that our computations satisfy this scaling, we considered the case $\Lambda$CDM0 (first column of Table \ref{tableCDMhist}), varying the amplitude from $A_{s1} = 1.885 \cdot 10^{-9}$ to $A_{s2} = 2.011 \cdot 10^{-9}$. The ratio of the two amplitudes\footnote{For these choices, the corresponding values of $\sigma_{8}$ at $z = 0$ are compatible within $\simeq 1 \sigma$ with available constraints. From the simulation, we got $\sigma_{8} = 0.804$ and $\sigma_{8} = 0.831$, respectively.} is $A_{s1}/A_{s2} = 0.93735$, 
while 
$\mathcal{h} \sigma^{2}_{1}(z)/\sigma^{2}_{2}(z) \mathcal{i}  = 0.93742$. 
This test confirms the expected general scaling law and the good accuracy level achieved by the numerical integration.
 
\subsubsection{Variable cut-off parameter}

We simply extend here the results found in the previous section, allowing for a variable cut-off parameter. 
We analyzed the cases: $k_{max} = 200$, $400$, $600$, $800$ and $1000$. 
Combining the results found in Sects. \ref{clump_supp} and \ref{constcut_varcosmo}, we find the following fitting formulas:

\begin{equation}
\begin{cases}
\sigma^{2}_{mod}(k_{ref}) = \sigma^{2}_{fit}(k_{ref}) \cdot  \sigma_{corr} \\\\
\sigma^{2}_{mod}(k) = \sigma^{2}_{mod}(k_{ref}) / \sigma^{2}_{ratio}(z)
\label{modvar}
\end{cases} ,
\end{equation}

\noindent where $\sigma^{2}_{mod}(k)$ and $\sigma^{2}_{fit}(k_{ref})$ are respectively the variances of the generic model and of the reference one, and $\sigma^{2}_{ratio}(z)$ is the fitting function of Eq.\;(\ref{fitratio}). 

\begin{figure}
\centering
\includegraphics[scale=0.5]{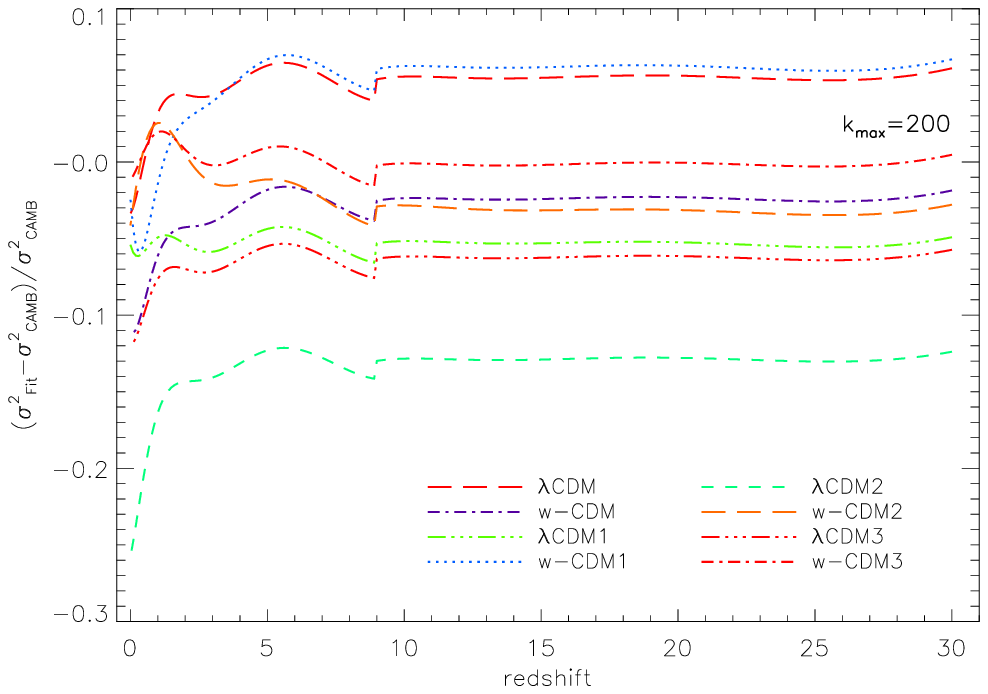}
\caption{Relative error of the variance.} 
\label{fig:err200}
\end{figure}

\begin{figure}
\centering
\includegraphics[scale=0.5]{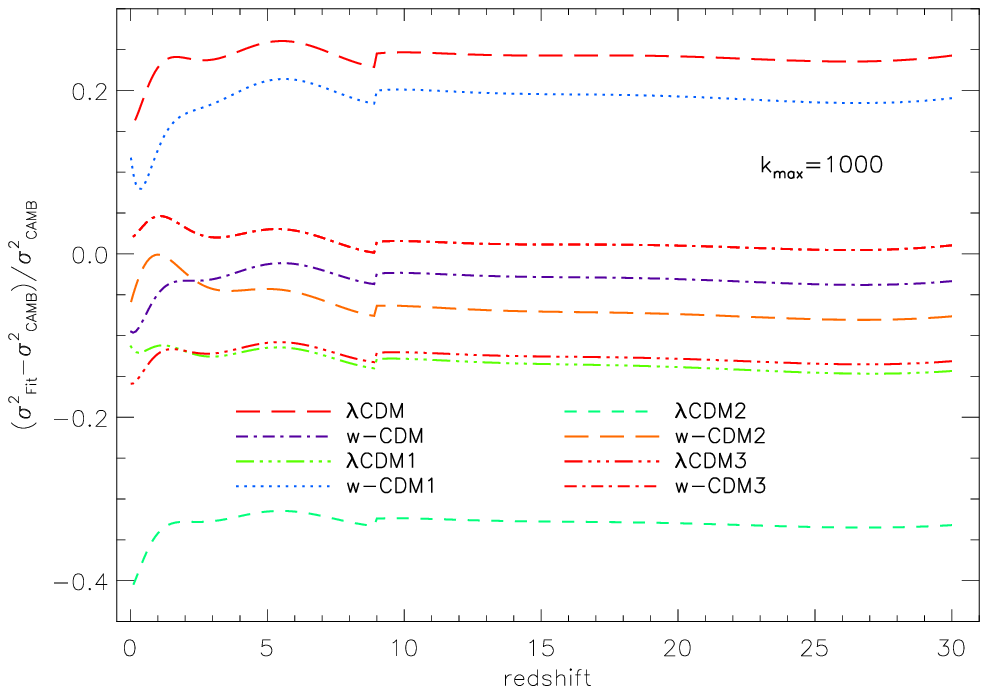}
\caption{Relative error of the variance.}  
\label{fig:err1000}
\end{figure}

Figs.\;\ref{fig:err200} and \ref{fig:err1000} show the relative errors of the variance for two values of the cut-off wavenumber. 
As expected, the found approximations are more accurate for a cut-off parameter closer to the reference one. 
In spite of the wide range explored for the cut-off parameter, the relative errors of these simple formulas are always within $\simeq 20\%$ for all scenarios, but for the $\Lambda$CDM2 prescription, for which the error is about $40\%$ for $k_{max}=1000$. Finally, Fig. \ref{fig:err1000} shows that the relative error, $E_{r}$, of the above fitting formulas is not strongly dependent on redshift. Therefore, even for high values of the cut-off parameter, a simple multiplicative factor $1/(1+E_{r})$ allows to correct the above formulas keeping the relative error within $\sim 10-15$\%.

\subsection{\textbf{\emph{Planck}} clumping factor} 
\label{planck}

\begin{figure}
\centering
\includegraphics[scale=0.5]{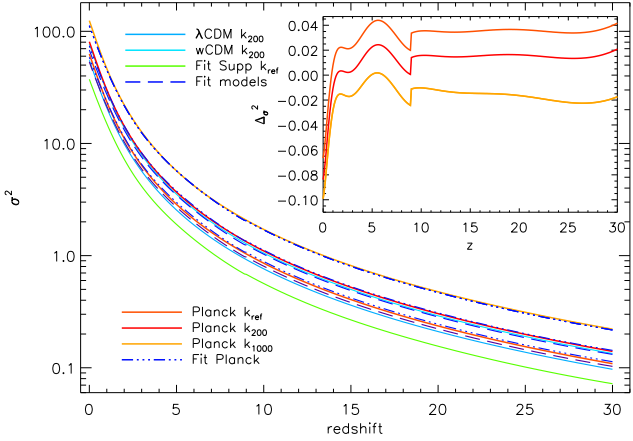}
\caption{Variance for the {\it Planck} cosmology (solid lines) compared to the approximated case (dash dot lines) and to alternative cosmological models (top curves in legend). The inset shows the corresponding relative error, only for the {\it Planck} estimate.}
\label{fig:sgmP}
\end{figure}

Given the recent delivery of the first {\it Planck} cosmological data and results, it is interesting to derive suitable approximations of the clumping factor for cosmological parameters in agreement with those found 
by the {\it Planck} Collaboration. To this aim, we assumed the {\it Planck} cosmological parameters set and derived the variance (see Fig. \ref{fig:sgmP}) using the standard CAMB. In this analysis, we took into account the cases $k_{max}$ = 100, 200, 1000, i.e. the reference starting point and some remarkable cases. We plot also for comparison the results found before for few other cosmologies. In the plot, the ``Fit models'' curves (dashed lines) refer to the variance of these scenarios numerically computed, while the ``Fit Planck'' curves (dashed dotted lines) describe the approximations for $\sigma^{2} (z)$ found assuming {\it Planck} cosmological parameters. We found that they can be relatively well represented by the formulas in Sect. \ref{clump_supp}, but corrected with a multiplicative factor 
$\sigma_{corr} = 1.57$. %$\sigma_{corr} = 1.571428$. 
In the figure, the inset shows the relative error in this approximation. A unique correction factor leads to a certain underestimation (overestimation) of $\sigma^{2} (z)$ for $k_{max}=1000$ (resp. for the reference cut-off wavenumber $k_{max}=100$), but the found discrepancy is typically $\lsim 4\%$. 
 
\section{Free-free distortion}
\label{FF}

To estimate the spectral distortions produced by the free-free process and amplified by the clumping, we developed a dedicated Fotran90 tool, which consists of different modules. The output of the code, the free-free parameter $y_{B}$ of Eq. (\ref{eq:free}) has been evaluated in the redshift interval [0, 30] with the D01AJF routine of the NAG libraries, based on the Gaunt 10-point and Kronrod 21-point rules. At the same time, we tested the accuracy of the free Fortran Numerical Recipes (NR) libraries \citep{press92} against the NAG routines, by implementing the code also with the NR Gaussian quadrature.

\begin{figure}
\centering
\includegraphics[scale=0.5]{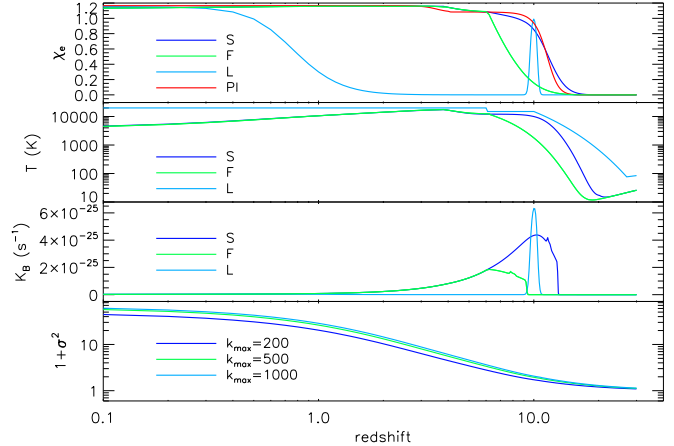}
\caption{Redshift dependence of the electron ionization fraction (top panel), the electron temperature (top middle panel) and the bremsstrahlung rate (bottom middle panel) derived for the astrophysical and phenomenological reionization histories analyzed in this study. The bottom panel represent the clumping factor for three values of $k_{max}$, being in this approximation independent of the assumed reionization history. }
\label{fig:reion}
\end{figure}

In Fig. \ref{fig:reion}, we compare the time varying ionization fraction for all the scenarios accounted in the study (top panel), and the evolution of the electron temperature (top middle panel), of the bremsstrahlung rate (bottom middle panel) and, lastly, of the clumping factor (bottom panel) for S, F and L models. 
From the first panel we can see how the {\it Planck} cosmology (at the basis of the standard CAMB) give rise to a ionization fraction which is highly comparable with the S model. Concerning the second panel, since CAMB assumes a mathematical representation for the evolution of the cosmic plasma in order to track the CMB anisotropies, we were not able to derive, at a glance, the electron temperature, variable required in the Saha equations. Thus, since $\tau_{F} < \tau_{Pl} < \tau_{S}$, we hypothesized that $T^{Pl}_{e}$ history could have been between two limits, the F and S prescriptions. Thereby, we evaluated the spectral distortions produced with {\it Planck} data in the two cases.

Globally, as expected, the clumping factor gives an important contribution at low redshift, when the plasma is characterized by a high ionization fraction, as highlighted by the comparison of the panels.

The CMB spectrum, in terms of the brightness temperature, can be 
described by the relation:

\begin{equation}
T_{br}(x) \simeq \left ( \frac{y_{B}(x)}{x^{2}} - 2 u \phi_{i} + \phi_{i} \right ) T_{r} \; ,
\label{eq:tbr}
\end{equation}
 
\noindent which depends on the frequency only (and holds at any redshift, provided that $y_{B}$ and $u$ are integrated over the corresponding interval $[z,z_{h}]$).

%\newpage

\section{Results}
\label{results}

\begin{figure}
\centering
\includegraphics[scale=0.5]{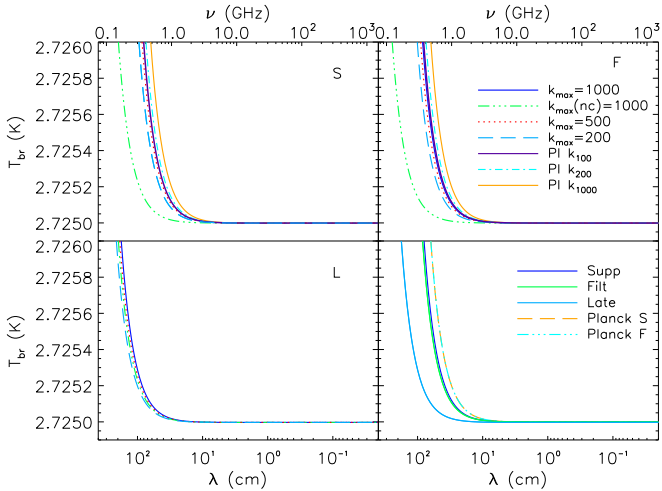} %FFandClump.ps
\caption{CMB spectral distortions induced by free-free for the three different reionization histories (see text for details) and for three values of $k_{max}$. The curve corresponding to $k_{max} (nc) = 1000$ (green dashed-triple dotted line) neglects the clumping factor, and is displayed in all panels for comparison with the case $k_{max = 1000}$. See also the text.}
\label{fig:FFclump}
\end{figure}

The global free-free spectral distortions are shown in Fig. \ref{fig:FFclump}, where each panel displays one of the possible scenarios accounted in the study, i.e. suppression (S), filtering (F) and late double peaked (L), and for three independent maximum wave-numbers $k_{max}$. We also report a curve, $k_{max}(nc)$, in which we do not account for the clumping amplification factor, aiming at directly unravel the impact of a primordial IGM density contrast on CMB spectral distortions. 

The free-free distortions induced by the {\it Planck} cosmology, as reported in 'S' and 'F' panels of figure, has been evaluated adopting the astrophysical electron temperatures for the numerical solution of the Saha equations. Indeed, since the standard CAMB does not track the evolution of the electron temperature, providing the ionization fraction as a rough parametrization which maps the fraction into the recombination residuals during the matter dominated era, we assumed the suppression and filtering electron temperature as upper and lower limits. The reason for this choice resides in their corresponding reionization optical depth values, for which $\tau_{{\it Planck}} = 0.0949$ lies almost in the middle of the models optical depth. The bottom panel in figure shows almost indistinguishable curves (dashed and dot-dashed lines) meaning that the spectral distortions generated by the two methods are significantly comparable, thus confirming the independence of this kind of distortion from the ionization history. 

The entity of the distortions is, as expected, stronger in the case of the astrophysical models, where the cosmic plasma becomes fully ionized since redshift $z \simeq 10$, in agreement with an increasing clumping factor, in comparison with the phenomenological scenario induced distortions (see Fig. \ref{fig:reion}). 
Also, the two astrophysical models predict signal with relatively low differences, much more smaller when clumping is included, since it introduces the most relevant amplification at $z \lsim 6$ when the two scenarios predict almost complete ionization.
Differently, the late history, in spite of being described by the same optical depth of the suppression model, is characterized by a first ionization peak taking place at $z \simeq 10$, where the clumping factor is almost negligible, followed by a rapid ionization decrease resulting into an almost neutral medium till recent epochs ($z \simeq 1$), and then by a significant ionization increase only at low redshift when the clumping factor significantly increases. 
For this reason, in the phenomenological model the effective outcome of the clumping is much less outstanding,
although we can again appreciate a tiny deviation of the CMB temperature between the case with (solid blue line) and without (dashed-triple-dotted green line) the inclusion of the clumping factor for an equivalent value of $k_{max}$. Finally, last panel in figure makes a comparison between the spectral distortions induced by these histories, reporting the (non preferential) case $k_{max} = 1000$. From this plot, it is easy to remark that the IGM density contrast noticeably affects the CMB brightness temperature evolution especially at decimeter wavelengths.

\begin{figure}
\centering
\includegraphics[scale=0.5]{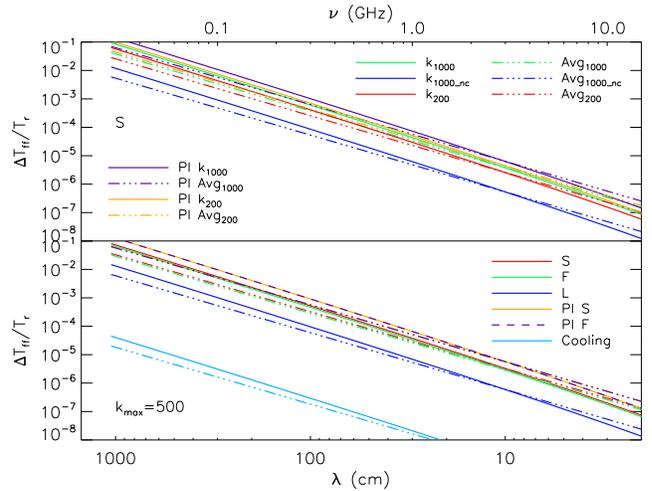}
\caption{Relative variation of the brightness temperature as function of the wavelength assuming the $y_{B} (x)$ value properly evaluated at each frequency in the entire range, and a value $\mathcal{h} y_{B}(x) \mathcal{i}$ averaged in the interval [1,100] cm, a relevant observational range for this process (lines denoted with Avg in legend). 
Upper panel: suppression model for $k_{max} = 1000$ (green line), $k_{max} = 200$ (red line) and $k_{max} = 1000$ without the correction due to the IGM density contrast (blue line). We show also the signal computed in the  case of {\it Planck} coupled with the thermal history of the suppression model.
Bottom panel: comparison between the three reionization histories for $k_{max} = 500$ and the case of {\it Planck} coupled with the thermal history of the suppression and filtering model, including clumping. 
We report also an approximate evaluation of the amplitude of the (negative) free-free signal induced by cooling in the case of Bose-Einstein condensation. See also the text.}
\label{fig:FFdiff}
\end{figure}

Basically, in order to describe the effect induced by free-free mechanism in terms of temperature excess at different frequencies, we can rewrite Eq. (\ref{eq:tbr}) as:

\begin{equation}
\frac{\Delta T_{ff}}{T_{r}}(x) \simeq \frac{y_{B}(x)}{x^{2}}  \; ,
\label{eq:tff}
\end{equation}

\noindent where we have defined $\Delta T_{ff} = (T_{br} - T_{r} \phi_{i})$. This is illustrated in Fig. \ref{fig:FFdiff} where the lines denoted with the term Avg$_{k_{max}}$ refer to a temperature variation derived assuming a value of $y_{B}$ averaged over a suitable wavelength range. 

In all cases, the steeper lines refer to the complete computation, while the flatter lines are derived using the averaged $\mathcal{h} y_{B}(x) \mathcal{i}$ (for which the exact value depends also on the considered frequency range) and thus show the simple wavelength dependence $\propto \lambda^{2}$. The slope derived including all the effects is $\sim 0.2$ steeper than that derived using $\mathcal{h} y_{B}(x) \mathcal{i}$.

We finally point out that, although a power law with a spectral index slightly larger than 2 represents a certain improvement with respect to the simple assumption of constant $y_{B}(\lambda)$, a dependence $y_{B}(\lambda) \simeq a \, {\log} \lambda + b$ well approximates $y_{B}(\lambda)$ at $\lambda \gsim 1.5$\,cm for all the considered models, but with slightly different values of $a$ and $b$. 
This behavior simply reflects the bremsstrahlung Gaunt factor dependence on $\ln (x_{e})$ at low frequencies, but with coefficients $a$ and $b$ globally depending on the selected power spectrum cutoff parameter and reionization scenario, in the relevant redshift range. Appendix \ref{ybvalues} reports the values of $y_{B}(\lambda)$ computed for the considered models at some representative wavelengths. 
From the values at the two longest wavelengths we derived the above parameters $a$ and $b$ and their dependence on $k_{max}$ that allow to find the relation  
$y_{B}(\lambda) \simeq  A k_{max}^{m} ({\log} \lambda + B)$, with $A$ and $B$ respectively dependent and independent of the model. 
Remarkably, $y_{B}(\lambda)$ can be approximated by a simple power law dependence on $k_{max}$ with a slope, $m$, weakly dependent on the model.

\begin{figure}
\centering
\includegraphics[scale=0.5]{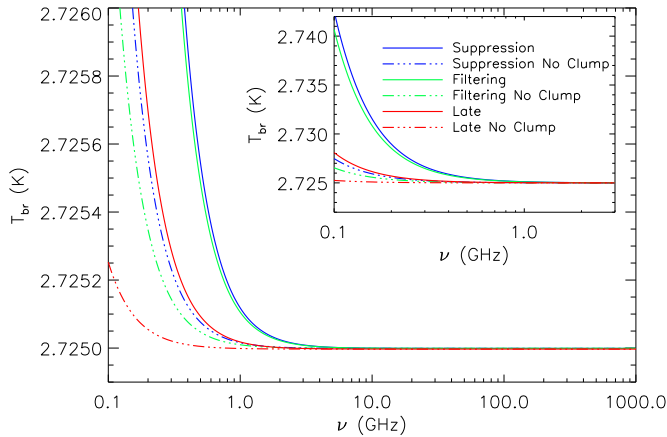}
\caption{Free-free distortion in terms of brightness temperature computed from combining Boltzmann and free-free CMB distortion numerical codes produced by the two astrophysical reionization histories and a phenomenological model at $k_{max}=1000$. CMB temperature distortions induced by free-free accounting for the clumping factor are found to be approximately 10 times larger than the ones derived neglecting IGM density contrast. The inset allows to appreciate the excess at low frequencies.}
\label{fig:FFska}
\end{figure}

\section{Conclusion}
\label{conclu}

We have developed a method able to characterize the variance of the matter power spectrum for different cosmologies and different values of the cut-off parameter $k_{max}$, the maximum wavenumber adopted to integrate the power spectrum for the variance computation, exploiting the output of the Boltzmann code CAMB. These numerical results and the found approximations may be used for a wide set of cosmological applications. 
An interesting example is the investigation of the reionization process possibly driven at high redshifts ($[5,10]$) by galaxy populations with active star formation, as recently explored in the model by \cite{salvaterra_etal_11} where the recombination rate density in the IGM is proportional to the HII clumping factor \citep{madau99}, a parameter substantially identical to that studied in this work. 
The suitable approximations presented here could be implemented in such formalism to improve the precision and the astrophysical motivations of the predictions.

The main scope of this paper is the evaluation of the efficiency of the bremsstrahlung process and the free-free distortion of the CMB spectrum when the density contrast of the IGM in an inhomogeneous Universe is taken into account. We developed a numerical code for the evaluation of the free-free distortion parameter and applied it to some reionization histories well defined in literature: two astrophysical models, suppression and filtering; a phenomenological description, late double peaked; a ionization history compatible with the recent derivation of cosmological parameters by the {\it Planck} mission. In the code, we implemented a routine aimed at finding a numerical solution of the Saha equations for a variable mixture of hydrogen and helium primordial abundances, providing a precise evaluation of their corresponding ionization states. We also performed a dedicated routine to weight on their Gaunt factors. 

We showed, in a redshift interval $z \in [0,30]$, the electron ionization fraction and the electron temperature for the histories accounted in the code, the corresponding bremsstrahlung rate and the clumping, the correction factor to the free-free term which accounts for the IGM density contrast. We compare the effect on free-free distortion coming from different values of $k_{max}$ for the same reionization history and for different scenarios. 
Focussing on the wavelength dependence on the free-free excess, we compare the signal slope derived from the complete computation and assuming an $y_{B} (x)$ value averaged over a suitable wavelength range (from 1 cm to 1 m), obtaining a somewhat larger slope in the former case than in the latter.

The two astrophysical histories show much less relative different free-free distortions when the clumping factor is included, because an almost full ionization state is achieved in both cases when the clumping factor significantly amplifies the signal.
The late double peaked model turns out to be characterized by a significantly smaller free-free distortion, in spite of the assumption of a Thomson optical depth equal to that of the suppression model. This is related to the peculiar ionization fraction history, from a fully ionized medium at $z  \sim 10$ to an almost nearly neutral phase between 
$1 \lsim  z \lsim 10$ and then again to a fully ionization only at lower redshifts where the clumping factor becomes significant. 

\section{Discussion}
\label{discussion}

It is interesting to compare our results with those found in \cite{ponenteetal}, where the average free-free emission is obtained from $N$-body simulations 
(see also their discussion about limitations of numerical simulations and semi-analytical treatments, given the complexity of the astrophysical processes involved). 
The authors computed free-free fluxes projected along the line of sight in slices of pixelized 2D map and then derived the global signal adding over the slices.
In particular, we note that the results shown in Fig. \ref{fig:FFska} for the astrophysical models well agree with the level of distortion found by \cite{ponenteetal} when the free-free emissivity has been computed with the temperature of gas particles derived from the simulation (compare with the solid line in their figure 5). 

\begin{figure}
\centering
\includegraphics[scale=0.5]{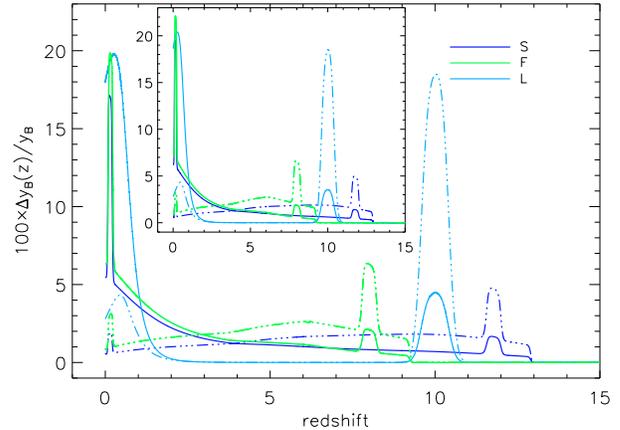}
\caption{Redshift dependence of the partial contribution to the free-free parameter, evaluated per redshift bin $\Delta z = 0.2$ at each $z$ for $k_{max} = 100$ (and $k_{max} = 500$ in the inset). 
Being normalized to the global value of $y_{B}$ at each wavelength, $\Delta y_{B} (z)/ y_{B}$ turns to be essentially independent of the wavelength, 
thus we present only the result at 10 cm. 
The dashed-dotted lines display the results found neglecting the clumping, for the corresponding histories reported in legend.}
\label{fig:clump}
\end{figure}

In order to identify the epochs giving the major contributions to $y_{B}$, where it will be particularly relevant to focus observational and theoretical efforts, 
we consider a redshift bin $\Delta z = 0.2$ and evaluated the partial contribution of each bin to $y_{B}(\lambda)$ for some characteristic wavelengths. 
% ($\lambda$ = 0.1, 0.3, 1, 3, 10, 30, 100 cm). 
Each scenario has been investigated with and without the inclusion of clumping factor. As evident from Fig. \ref{fig:clump}, the IGM clumping factor largely affects the evolution of the free-free term, particularly at low redshift where the contribution increases of about an order of magnitude. The redshift dependence of $\Delta y_{B} (z)/ y_{B}$ shows two epochs of particular relevance, as a consequence of the behaviors displayed in Fig. \ref{fig:reion}. The first occurs at high redshift when the ionization fraction significantly raises or, in the case of the L history, at the redshift corresponding to the peak in the ionization fraction. The second occurs at low redshift when full ionization is achieved and the clumping factor becomes particularly high. 

Fig. \ref{fig:clump} explains also the (weak) dependence on the model of the slope, $m$, characterizing the power law dependence of $y_{B}(\lambda)$ on the cut-off wavenumber $k_{max}$: $m$ increases with the 
increase of $\Delta y_{B} (z)/ y_{B}$ at low redshift, as a result of the larger relevance of clumping. When the astrophysical scenario will be well understood, free-free signatures could be in principle used to further test dark matter properties through their effect on the power spectrum at small scales.

In principle, a complete analysis of the problem requires the proper treatment of the temperature (spatial) fluctuations of the ionized IGM, as mentioned in Sect. \ref{brem}.
To clarify the possible relevance of this aspect, it is interesting to compare the terms 
$K_{B}/x_{e}^{3} \propto \phi^{-1/2}$ and $e^{-x_{e}} \simeq 1-x_{e}$ estimated adopting an averaged IGM temperature or averaged over a representative volume. 
When the contribution to free-free distortion is remarkable, the electron temperature is significantly larger than the CMB temperature, i.e. $\phi \gg 1$. 
For example, even at $\sim 1$\,cm and for $\phi \sim 10$ a temperature fluctuation of $\sim 30$\% implies a change of only $\sim 1.6$\% in the term $1-x_{e}$ and the effect clearly decreases at increasing temperature as well as at longer wavelengths where free-free distortion is more important.
Let us consider the term $K_{B}/x_{e}^{3} \propto \phi^{-1/2}$. The ratio $\langle \phi^{-1/2} \rangle /  \bar{\phi}^{-1/2}$, where $\bar{\phi} = \langle \phi \rangle$, obviously depends on the level of IGM temperature fluctuations, 
$\delta \phi$, and on the shape of their distribution function. In general, to second order in $\delta \phi$,
$\langle \phi^{-1/2} \rangle /  \bar{\phi}^{-1/2}  = 1+ (3/8) (\sigma_{\phi}/\bar{\phi})^{2}$\,, where $\sigma_{\phi}^{2} = \langle (\delta \phi)^2 \rangle$ is the IGM temperature fluctuation variance, while in the particular case of a Gaussian distribution of $\delta \phi$ one gets $\langle \phi^{-1/2} \rangle /  \bar{\phi}^{-1/2}  = 1+ (3/8) (\sigma_{\phi}/\bar{\phi})^{2} + (105/128) (\sigma_{\phi}/\bar{\phi})^{4} + (3465/1024) (\sigma_{\phi}/\bar{\phi})^{6} + ...$\,. 
A compilation of temperature evaluations through hydrodynamical simulations and line-of-sight radiative transfer approaches is presented in \cite{boltonetal2010} and estimates of the IGM temperature from a semi-numerical model are given by \cite{raskuttietal2012} for some representative redshift values and intervals. The authors also compared theoretical predictions with direct measurement of the IGM temperature around quasars. 
Temperature standard deviations are quoted to be always less than $\simeq 50$\% and typically less than $\simeq30$\%. Also, \cite{ciardietal2012} reported IGM volume averaged temperature for various 
ionizing emissivity models, focussing on a redshift range $z=[6,14]$. Again, the relative differences between the considered model predictions are less than $\simeq 30$\%. Therefore, even assuming as a generous upper limit for the whole redshift range relevant for free-free distortion $\sigma_{\phi}/\bar{\phi} \simeq 30$\%, we find $\langle \phi^{-1/2} \rangle /  \bar{\phi}^{-1/2} - 1 \simeq 4$\%. 

In general, the bremsstrahlung rate involves a product proportional to $\rho^2 \phi^{-1/2}$. We should then include in the treatment also the mixing between density and temperature fluctuations. By expanding in Taylor's series, one can see that a further second order term, not  included in the previous discussion, appears: $- \langle  \bar{\rho} \bar{\phi}^{-3/2} \delta \rho \delta \phi \rangle$. Its precise evaluation requires the treatment of possible physical correlations between the evolution of density and temperature fluctuations. 
We can write $\delta \rho \delta \phi$ as the sum of an uncorrelated and a correlated term, $\delta \rho \delta \phi =  (\delta \rho \delta \phi)_{u}  + (\delta \rho \delta \phi)_{c}$. The average of the uncorrelated term clearly vanishes. We are then driven to consider the term
$- \bar{\rho} \bar{\phi}^{-3/2} \langle (\delta \rho \delta \phi)_{c} \rangle$ that, divided by the homogeneous term $\bar{\rho}^{2} \bar{\phi}^{-1/2}$, introduces the further correction term
$ - (\sigma_{\rho} / \bar{\rho}) (\sigma_{\phi} / \bar{\phi}) \langle [(\delta \rho / \sigma_{\rho}) (\delta \phi / \sigma_{\phi} )]_{c}  \rangle$ to be included in the global correction factor\footnote{It could weakly reduce or amplify the free-free signal according to the dominance of correlation or anticorrelation of $\delta \rho$ and $\delta \phi$, respectively.}.
%Clearly, $(\sigma_{\rho} / \bar{\rho}) = (1+\sigma^{2})^{1/2}$ 
Clearly, $(\sigma_{\rho} / \bar{\rho}) = \sigma$ is much smaller than the main clumping correction term, $(1+\sigma^{2})$,
object of this work, while $(\sigma_{\phi} / \bar{\phi}) \lsim 0.3$. 
Therefore, unless the average of the correlated density-temperature fluctuations is much larger than the product of their root mean squares, the mixing term is subleading. 

We then deduce that IGM temperature fluctuations could likely introduce only a very small modification of the global free-free distortion.
 
\subsection{Comparison with other spectral distortions and future observational perspectives}

Within this context, it is important to compare the amplitude of free-free distortion predicted in these models with that expected from other kinds of spectral distortions. We consider here few examples of unavoidable spectral distortions, in particular the Compton scattering between CMB photons and electrons. Actually, the reionization process produce an electron heating which cause a distortion proportional to the fractional amount of energy exchanged during the interaction, the so-called Comptonization parameter $u \simeq (1/4) \Delta\varepsilon/\varepsilon_{i}$.

For the astrophysical reionization models considered here, the parameter is $u \sim (0.965 - 1.69) \times 10^{-7}$ \citep{paperII}.
The ratio between the brightness temperature excess by free-free distortion and decrement by Comptonization can be approximated by:

\begin{equation}
{\Delta T_{ff} \over \Delta T_{C}} \simeq 1.79 \times 10^{-2} \left({\lambda \over {\rm cm}} \right)^{2} \left( { y_{B}/10^{-9} \over u/10^{-7} }\right) \, . 
\label{ratioFFC}
\end{equation}

\noindent
Therefore, assuming $y_{B} \sim 5 \times 10^{-9}$ (see Appendix \ref{ybvalues}), we find that the free-free distortion exceeds the Componization decrement for $\lambda \gsim 3-5$\,cm ($\nu \lsim 6-10$ \,GHz), below the frequency range proposed for both PIXIE \citep{kogutpixie} and PRISM\footnote{http://www.prism-mission.org/}, spanning from 30 GHz to 6 THz.

Other kinds of unavoidable spectral distortions are Bose-Einstein (BE) like distorted spectra produced by the dissipation of primordial perturbations at small scales
(\cite{Sunyaev70,Daly91,Hu94,Chluba12,2013JCAP...02..036P})
which produces a positive dimensionless chemical potential, $\mu_{0}$, and Bose-Einstein condensation of CMB by colder electrons (\cite{ChlubaSun12}; see also \cite{khatrietal2012, sunyaevkhatri2013} for recent reviews), which gives a negative chemical potential. The two kinds of distortions are characterized by an amplitude, respectively, in the 
range\footnote{Since very small scales not explored by current CMB anisotropy data are relevant in this context, a wide range of primordial spectral index needs to be exploited. A wider range of chemical potentials is found by \cite{chlubaal12} allowing also for variations of the amplitude of primordial perturbations at very small scales, as motivated by different inflation models.}
 $\sim 1.5 \times 10^{-9} - 10^{-7}$
(and in particular $\simeq 2.52 \times 10^{-8}$ for $n_{S}=0.96$, without running)
and $\simeq 3.08 \times 10^{-9}$. 
It is interesting to compare the free-free distortion with the Bose-Einstein (BE) like distorted spectrum in two particular regions: at the wavelength $\lambda_{m}$, dependent on $\hat{\Omega}_{b}$, where the modified BE spectrum shows the maximum distortion, $(\Delta T/T)_{m}$, and at $\lambda \lsim 3$\,cm where it is very weakly dependent on $\hat{\Omega}_{b}$. Adopting respectively the approximations by \cite{BuriDanZotAA91} for $\lambda_{m}$ and $(\Delta T/T)_{m}$ and the $x \gg \mu_{0}$ limit in the BE formula, we find
 
\begin{equation}
{\Delta T_{ff} \over \Delta T_{BE}} \simeq 91.0 \left({\hat{\Omega}_{b} \over 0.1}\right)^{-2/3} \left( { y_{B}/10^{-9} \over \mu_{0}/10^{-9} }\right) \, 
\label{ratioFFC1}
\end{equation}

\noindent
at $\lambda \simeq \lambda_{m}$, and

\begin{equation}
{\Delta T_{ff} \over \Delta T_{BE}} \simeq 1.89  \left({\lambda \over {\rm cm}} \right) \left( { y_{B}/10^{-9} \over \mu_{0}/10^{-9} }\right) \, 
\label{ratioFFC2}
\end{equation}

\noindent
at $\lambda \lsim 3$\,cm.

Therefore, at $\lambda \simeq \lambda_{m}$, the amplitude of free-free distortion always exceeds, or even overwhelms, that of BE-like distortion predicted in these models.
%(respectively we find $\sim 5-330$ or of $\sim 160$).
Instead, assuming as reference $y_{B} \sim 5 \times 10^{-9}$ and $\mu_{0} \simeq 2.52 \times 10^{-8}$, the predicted BE-like distortion exceeds the free-free excess for $\lambda \lsim 2.7$\,cm. 

To provide an estimate of negative free-free distortion generated by colder electrons in the case of Bose-Einstein condensation, we exploited the temperature evolution reported in \cite{ChlubaSun12} (see their figure 2), coupled with the ionization history computed with CAMB assuming cosmological parameters by {\it Planck} (see Sect. \ref{planck}). A certainly generous upper limit to $y_{B}$ amplitude, computed in different redshift ranges,
is obtained by Eq. (\ref{eq:free}) integrating over the interval $10 < z < 10^{5}$, before the IGM temperature increase by reionization, when clumping is negligible. 
The result is reported in Fig. \ref{fig:FFdiff}, showing that the distortion is two (or more) orders of magnitude less than free-free signal generated in any reionization model.

We finally compare the predicted free-free distortion with the cosmological signal expected from the HI 21-cm background in the same reionization scenarios. Looking at figure 5 of \cite{paperI} it is found to 
prevail over the free-free distortion only in a restricted frequency window between $\sim 90$ and $\sim 130$\,MHz or $\sim 110$ and $\sim 250$\,MHz for the suppression and filtering model, respectively. On the other hand, 
the two signals display very different frequency dependences that can be exploited to distinguish them.

These estimates imply that a future CMB experiment at millimeter wavelengths with a sensitivity to absolute temperature as in the PRISM
proposal \citep{PRISMWP}, having the ambitious goal of detecting primordial BE-like and later Comptonization distortions, will be likely affected only weakly by late free-free distortion contribution, at least for not extreme reionization models. Future accurate CMB spectrum observations at longer wavelengths, extending and 
improving the recent TRIS \citep{Gervasietal08} and ARCADE 2 results\footnote{http://asd.gsfc.nasa.gov/archive/arcade/} \citep{Singaletal11,Seiffertetal11}, and observations in the radio domain are particularly interesting for the detection of the cosmological reionization free-free distortion. The amplitude of the signal predicted in this work (and, obviously, in more extreme models such that considered for example by \cite{Oh99}) will be accessible to the sensitivity of the SKA\footnote{http://www.skatelescope.org/} \citep{SKAScience} and of the next-generation radio telescopes (e.g. LOFAR, MWA, ASKAP, MeerKAT). 
An accurate inter-frequency absolute calibration and a dedicated data analysis strategy able to estimate the background signal and its spectral shape in interferometric observations will be an important step to firmly detect and study this cosmological imprint.

\appendix
\section{Saha equations}
\label{sahaEq}

The process of reionization can be studied on the basis of the Saha equation that describes the ratio between 
two different ionization states of an element. Known the free electron number density and the temperature, we can write

\begin{equation}
\frac{n_{i+1}n_{e}}{n_{i}}=\frac{2}{\Lambda^{3}}\frac{g_{i+1}}{g_{i}}e^{{-\frac{\varepsilon_{i+1}-\varepsilon_{i}}{k_{B}T}}} \, ,
\end{equation}

\noindent
where $n_{i}$ is the density of atoms in the $i-th$ state of ionization, $n_{e}$ the electron density, 
$g_{i}$ the degeneracy of states for the $i-$ions, $\varepsilon_{i}$ the energy required to remove $i$ electrons from a neutral atom, 
and $\Lambda$ the thermal de Broglie wavelength of an electron, defined by

\begin{equation}
\Lambda=\sqrt{\frac{h^{2}}{2\pi m_{e}k_{B}T}} \, .
\end{equation}

%The unknowns $\chi_{H},\chi_{H^{+}},\chi_{He},\chi_{He^{+}},\chi_{He^{++}}$, 
%defined as the relative abundances of the different ionization states of each element with respect to the global number density of the element,
%can now be computed from this set of equation, separately for all species.

From the nuclei conservation law for $H$ and $He$ we also know that

\begin{eqnarray}
n_{H}^{tot} & = & \frac{\rho_{b}}{m_{b}} \left[ f_{H} \left( \chi_{H}+\chi_{H^{+}} \right) \right] \\ 
n_{He}^{tot} & = & \frac{\rho_{b}}{m_{b}} \left[ \frac{1-f_{H}}{4} \left( \chi_{He}+\chi_{He^{+}}+\chi_{He^{++}} \right) \right] \, ,
\end{eqnarray}

\noindent
where the sum of the ionization fraction of each specie is equal to unity.

\subsection{Hydrogen Ionization fraction}

In the case of $H$, taking into account the Saha equation and the charge conservation law, the system to be solved is:

\begin{equation}
\begin{cases}
\chi_{H} + \chi_{H^{+}} = 1 \\
\chi_{H^{+}} = \frac{2}{1+f_{H}} \frac{m_{b}}{\rho_{b}} \frac{\chi_{H}}{\chi_{e}} \frac{1}{{\Lambda^{3}}} e^{- \frac { \varepsilon_{H^{+}} - \varepsilon_{H}} { k_{B} T}} \; ,
\end{cases}
\end{equation}

\noindent being $2g_{H^{+}}/g_{H} = 1$. Defining $\Delta \varepsilon_{H} = \varepsilon_{H^{+}} - \varepsilon_{H}$ and

\begin{equation}
C_{1} = \frac{2}{1+f_{H}} \frac{m_{b}}{\rho_{b}} \frac{1} {\chi_{e} \Lambda^{3}} e^{- \frac { \Delta \varepsilon_{H} } { k_{B} T}} \, ,
\end{equation}

\noindent the solution is:

\begin{equation}
\begin{cases}
\chi_{H} = \frac{1}{1+C_{1}} \\  \\ 
\chi_{H^{+}} = \frac{C_{1}}{1+C_{1}} \, .
\end{cases}
\end{equation}

\subsection{Helium Ionization fraction}

For $He$, being the ratio $2g_{He^{+}}/g_{He} = 4 $ and $2g_{He^{++}}/g_{He^{+}} = 1$ the system is:

\begin{equation}
\begin{cases}
\chi_{He} + \chi_{He^{+}} + \chi_{He^{++}} = 1 \\
\chi_{He^{+}} = \frac{8}{1+f_{H}} \frac{m_{b}}{\rho_{b}} \frac{\chi_{He}}{\chi_{e}} \frac{1}{{\Lambda^{3}}} e^{- \frac { \varepsilon_{He^{+}} - \varepsilon_{He}} { k_{B} T}} \\ 
\chi_{He^{++}} = \frac{2}{1+f_{H}} \frac{m_{b}}{\rho_{b}} \frac{\chi_{He^{+}}}{\chi_{e}} \frac{1}{{\Lambda^{3}}} e^{- \frac { \varepsilon_{He^{++}} - \varepsilon_{He^{+}}} { k_{B} T}} \, .
\end{cases}
\end{equation}

Analogously to the hydrogen case, assuming $\Delta \varepsilon_{He} = \varepsilon_{He^{+}} - \varepsilon_{He}$ and $\Delta \varepsilon_{He^{+}} = \varepsilon_{He^{++}} - \varepsilon_{He^{+}}$:

\begin{eqnarray}
C_{2} & = & \frac{8}{1+f_{H}} \frac{m_{b}}{\rho_{b}} \frac{1} {\chi_{e} \Lambda^{3}} e^{- \frac { \Delta \varepsilon_{He^{+}} } { k_{B} T}} \\
C_{3} & = & \frac{2}{1+f_{H}} \frac{m_{b}}{\rho_{b}} \frac{1} {\chi_{e} \Lambda^{3}} e^{- \frac { \Delta \varepsilon_{H}e } { k_{B} T}} \, ,
\end{eqnarray}

\noindent the solution for the Helium is:

\begin{equation}
\begin{cases}
\chi_{He} = \frac{1}{1+ C_{2} + C_{2}C_{3}} \\\\
\chi_{He^{+}} = \frac{C_{2}}{1+ C_{2} + C_{2}C_{3}}\\\\
\chi_{He^{++}} = \frac{C_{2}C_{3}}{1+ C_{2} + C_{2}C_{3}} \, .
\end{cases}
\end{equation}

\section{Fit coefficients}
\label{impleprod}

The coefficients to be applied to Eqs. (\ref{eq:ab}) and (\ref{fitratio}) are:

\begin{eqnarray}
a_{0} & = & 0.51385 \;\;\;\;\;\;\; \;\;\;\;\;\;\; \;\;\;\; b_{0} = -8.9452\cdot 10^{-5} \nonumber \\
a_{1} & = & 0.85505 \;\;\;\;\;\;\;\;\;\;\;\;\;\; \;\;\;\;  b_{1} = 4.7704\cdot 10^{-5} \nonumber\\
a_{2} & = & 20.0 \;\;\;\;\;\;\;\;\;\;\;\;\;\;\;\;\;\;\;\;\; \;\;\; b_{2} = 20.0 \\
a_{3} & = & 26.377 \;\;\;\;\;\;\;\;\;\; \;\;\;\;\;\;\; \;\;\; b_{3} = 0.27323 \nonumber \\
a_{4} & = & 0.59601 \;\;\;\;\;\;\;\;\;\;\;\;\;\; \;\;\;\;\; b_{4} = 9.8741\cdot 10^{-5} \nonumber \\
a_{5} & = & 235.98 \;\;\;\;\;\;\;\;\;\;\;\;\;\;\;\;\; \;\;\; b_{5} = 1034.8 \nonumber
\end{eqnarray} 

\noindent
The coefficient obtained for the fit represented by Eq. (\ref{eqfit}) are:

\begin{eqnarray}
l_{0} & = & 0.27464 \;\;\;\;\;\;\; \;\;\;\;\;\;\; \;\;\;\;  h_{0} = 0.033389 \nonumber\\
l_{1} & = & 28.58 \;\;\;\;\;\;\;\ \;\;\;\;\;\;\; \;\;\;\;\;\;\;h_{1} = 0.24724 \nonumber\\
l_{2} & = & -1.574\cdot 10^{-13} \;\;\;\;\;\;\;\ h_{2} = 9.0 \\
l_{3} & = & 0.84439 \;\;\;\;\;\;\;\ \;\;\;\;\;\;\; \;\;\;\; h_{3} = 2.7819 \nonumber\\
l_{4} & = & 8.4836 \;\;\;\;\;\;\;\ \;\;\;\;\;\;\; \;\;\;\;\; h_{4} = 0.40036 \nonumber\\
l_{5} & = & 2.9378 \;\;\;\;\;\;\;\ \;\;\;\;\;\;\; \;\;\;\;\;h_{5} = 8.9756 \nonumber
\end{eqnarray} 

\section{{\emph{$y_{B}$}} values}
\label{ybvalues}

Table \ref{tabyb} reports the values of the free-free parameter at various wavelengths numerically computed for the suppression, filtering and late reionization models and for different cut-off parameters. 
The associated numerical error is always completely negligible in the case of quadrature performed with the NAG routine. The quadrature with NR is found to introduce a negligible underestimation 
(absolute relative error $\lsim {\rm few} \times 10^{-6}$).

The results at the 1.5 and 30 cm can be directly rewritten in terms of the parameters 
$a$ and $b$ characterizing $y_{B}(\lambda) \simeq a \, {\log} \lambda + b$ at $\lambda \gsim 1.5$\,cm. Finally, we found that $a$ and $b$ can approximated by a log-log linear dependence on $k_{max}$, i.e. $\log a \simeq m_{a} \, \log k_{max} + n_{a}$ and $\log b \simeq m_{b} \, \log k_{max} + n_{b}$, with $m_{b} \simeq m_{a} = m$, slightly dependent on the model, and
with $10^{n_{b}-n_{a}}$ independent of the model. Thus, $y_{B}(\lambda)$ can be approximated by a simple power law dependence on $k_{max}$. We find  
$y_{B}(\lambda) \simeq  A k_{max}^{m} ({\log} \lambda + B)$ with $B = 0.6288$ and $(A, m) = (1.337 \times 10^{-9}, 0.2275)$ for the S model (resp.
$(1.078 \times 10^{-9}, 0.2424)$ or $(2.096 \times 10^{-10} , 0.2482)$ for the F and L 
models)\footnote{This approximation has a relative error always $\lsim 3$\% for $k_{max} \gsim 100$, but tends to overestimate $y_{B}$ up to $\sim 14$\% at $k_{max} \lsim 20$.
Including in the fit also the results found at $k_{max} = 20$, we find the same value for $B$ but the coefficients $(A, m)$ slightly change to $(1.149 \times 10^{-9}, 0.2527)$, $(8.835 \times 10^{-10}, 0.2755)$, 
$(1.632 \times 10^{-10}, 0.2897)$ respectively for the S, F, L model, providing a fit relative error $\lsim 6$\%.}.

%$m, n_{a}, n_{b} \simeq 0.2275, -8.874, -9.075$ for the S model (resp.
%$0.2424, -8.967, -9.169$ or $0.2482, -9.679, -9.880$ for the F and L models). Thus, $y_{B}(\lambda)$ can be approximated by $y_{B}(\lambda) \simeq  k_{max}^{m} (10^{n_{a}} \, {\log} \lambda + 10^{n_{b}})$, 
%displaying a simple power law dependence on $k_{max}$.

\begin{table}
\centering
\begin{tabular}{| c | c | c | c | c | c |}
\hline
$\lambda$ (cm) & $0.1$ & $0.3$ & $1$ & $1.5$ & $30$ \\
\hline
$k_{100}^{s}$ & $1.293$ & $2.240$ & $2.971$ & $3.106$ & $8.128$ \\
$k_{200}^{s}$ & $1.462$ & $2.533$ & $3.360$ & $3.513$ & $9.192$ \\
$k_{500}^{s}$ & $1.860$ & $3.221$ & $4.273$ & $4.468$ & $11.69$ \\
$k_{1000}^{s}$ & $2.155$ & $3.733$ & $4.953$ & $5.178$ & $13.55$ \\
\hline
$k_{100}^{f}$ & $1.115$ & $1.932$ & $2.563$ & $2.680$ & $7.012$ \\
$k_{200}^{f}$ & $1.278$ & $2.214$ & $2.937$ & $3.071$ & $8.035$ \\
$k_{500}^{f}$ & $1.651$ & $2.859$ & $3.793$ & $3.965$ & $10.38$ \\
$k_{1000}^{f}$ & $1.924$ & $3.332$ & $4.420$ & $4.621$ & $12.09$ \\
\hline
$k_{100}^{l}$ & $0.218$ & $0.377$ & $0.500$ & $0.523$ & $1.369$ \\
$k_{200}^{l}$ & $0.262$ & $0.453$ & $0.601$ & $0.629$ & $1.645$ \\
$k_{500}^{l}$ & $0.340$ & $0.589$ & $0.781$ & $0.817$ & $2.137$ \\
$k_{1000}^{l}$ & $0.380$ & $0.660$ & $0.874$ & $0.914$ & $2.392$ \\
\hline
\end{tabular} 
\caption{Values of the free-free parameter $y_{B} (\lambda) / 10^{-9} $ for some
characteristic wavelengths; $\lambda = 1.5$ cm is the minimum wavelength from
which the relation between $y_{B}$ and $\log (\lambda)$ is linear.}
\label{tabyb}
\end{table}

\section*{Acknowledgements}

\noindent
We acknowledge support by ASI through ASI/INAF Agreement I/072/09/0 for
the {\it Planck} LFI Activity of Phase E2 and by MIUR through PRIN 2009. It is a pleasure to thank the anonymous referee for constructive comments.

\end{document}